# Broad-spectrum antiviral database (BSA-DB): a curated collection of broad-spectrum antiviral agents assembled by large language model mining and expert curation

Hovhannes Matevosyan[1,2], Narek Petrosyan[1,2], Levon Kharatyan[1,2], Iren Okminyan[2], Arman Kharatyan[2], Astghik Hakobyan[1], Vardan Harutyunyan[2,3], Mher Matevosyan[2,3], Hovakim Zakaryan[1,2,3,*]

1 Laboratory of Antiviral Drug Discovery, Institute of Molecular Biology of NAS, 0014, Yerevan, Armenia
2 Denovo Sciences Inc., 0060, Yerevan, Armenia.
3 Denovo Sciences Doha QFC Branch, Qatar Science and Technology Park, 210000, Doha, Qatar.

*Corresponding author email: hovakimz@denovosciences.ai; h_zakaryan@mb.sci.am

H.M. and N.P. contributed equally to this work. Author order was determined by the order in which the authors began working on the project.

## Abstract

Viruses included in the World Health Organization R&D Blueprint priority list pose a continuing epidemic and pandemic threat, highlighting the need for rapidly deployable antiviral discovery resources. Broad-spectrum antivirals (BSAs), which are active against multiple viruses, represent promising starting points for pandemic preparedness, yet relevant bioactivity data remain fragmented across public databases. Here we present BSA-DB, a freely accessible resource comprising 13,194 compounds with reported activity ≤ 10000 nM against at least two of 22 Blueprint viruses spanning 10 virus families (initially 23 viruses were selected, but Hantaan virus yielded no compounds meeting this criterion). BSA-DB was obtained from ChEMBL, PubChem and BindingDB, followed by systematic data curation and standardization of viral annotations. Cheminformatic analysis shows that the collection is chemically diverse rather than dominated by a few known antiviral chemotypes, and identifies recurrent scaffolds, structural clusters with broad virus coverage, and promising starting points for antiviral drug discovery. BSA-DB is available at https://bsadb.denovosciences.ai/ through an interactive web platform with search, filtering, visualization and bulk-download functionality, and will be continuously expanded with newly available and community-contributed data.

## 1. Introduction

The development of antiviral drugs is a lengthy and costly process (Adamson et al., 2021). Although some pharmaceutical companies continue to show interest in developing new antiviral therapeutics, this interest is often limited, and many efforts to bring effective medications to market remain incomplete. For example, mosnodenvir developed against the dengue virus (JNJ-1802; Phase 2 field study: NCT05201794) was discontinued for various reasons, despite promising preclinical and early clinical results (Durbin et al., 2025). Therefore, in light of limited commercial interest in virus-specific antiviral programs, a shift toward therapeutics with broad-spectrum antiviral (BSA) activity may represent a more sustainable approach to disease management (Martin et al., 2025). BSAs can be deployed rapidly against newly emerging or re-emerging viruses that exploit conserved host pathways or share structural homologies. This rapid deployability provides a critical early defense during outbreaks, when no virus-specific therapeutic or vaccine exists and when even modest antiviral efficacy can reduce morbidity, mortality, and transmission. Ribavirin provides a good example of a clinically approved BSA against respiratory syncytial virus and chronic hepatitis C virus that has been used to treat severe Lassa fever in the endemic region, highlighting the translational value of BSAs in urgent clinical contexts (Raabe et al., 2017).

While most BSAs have been discovered serendipitously, the rational design of BSAs could substantially increase the likelihood of identifying effective therapeutics. To advance BSA-related research, we developed a comprehensive database of existing BSAs curated from publicly available sources, such as PubChem (Kim et al., 2025), ChEMBL (Zdrazil et al., 2024), and BindingDB (Liu et al., 2007). We focused specifically on WHO-prioritized viruses (World Health Organization, 2024), extracting antiviral agents tested against 23 viruses across 11 viral families. Because public sources contain a vast number of entries for these viruses, we implemented a data-mining workflow based on a large language model (LLM) to efficiently identify relevant entries, retrieve associated experimental data, and integrate information from multiple sources. The resulting Broad-Spectrum Antiviral Database (BSA-DB) contains 13,194 compounds with reported activity against two or more viruses, either within the same viral family or across different families. Additionally, we conducted cheminformatics analyses to characterize chemical signatures, structural patterns, and promising scaffolds that may facilitate the discovery of new BSA candidates. The database is accessible through a user-friendly web interface.

## 2. Materials and Methods

### 2.1. Virus selection and data sources

Twenty-three viruses at high risk of causing Public Health Emergencies of International Concern were selected from the WHO R&D Blueprint. The full list and the keyword-based retrieval of data from PubChem, ChEMBL (release 36) and BindingDB (release 202601) are described in Supplementary Method 1 and Supplementary Note 5.

## 2.2. LLM-based extraction and validation

Where the viral target was not explicitly annotated, the assay name, title and abstract were processed with the DeepSeek-V3 API (DeepSeek-AI, 2024) using a task-specific prompt (Supplementary Note 6) to extract Target Name, Organism and Assay Goal Target. Extraction was benchmarked against expert-curated ChEMBL organism annotations and quantified as extraction accuracy; reproducibility was assessed by submitting 1,000 PubChem assays five times and scoring per-field consistency by fuzzy string matching. Prompts, formulas and per-virus results are given in Supplementary Method 2 and Supplementary Notes 1, 6 and 7.

## 2.3. Dataset assembly

Records were merged, SMILES were canonicalized with RDKit (version 2025.09.1), duplicates were removed, and activity values were converted to nanomolar units. Compounds with activity values that could not be converted to nM (percent inhibition, “Active”, “Potent”, etc.) were removed, and those with any value above 10,000 nM were excluded. We distinguish inclusion from potency throughout: the 10 µM (10,000 nM) value is an inclusion threshold defining database membership (compounds meeting it are “active”), whereas all tighter cut-offs applied to included compounds (1 µM, 100 nM, 10 nM) are potency thresholds, for which the terms “potent”/”highly potent” are reserved. Broad-spectrum activity was defined as activity (≤ 10,000 nM) against at least two priority viruses, and each compound was additionally annotated with the number of distinct viral families targeted to distinguish cross-family breadth from activity against multiple viruses of the same family. Of all compounds meeting the ≥ 2-virus criterion, 95% (12,510/13,194) were active across at least two viral families, while 5% (684) were active within a single family only. This threshold is the conventional boundary for assembling bioactivity collections from public repositories and for defining confirmed screening hits (Brideau et al., 2003; Koutsoukas et al., 2013; Mervin et al., 2015); it delimits the scope of the collection rather than defining lead-like potency, and stricter criteria are applied where compounds are prioritised (Sections 3.6 and 3.7). Each activity record was annotated with two controlled fields. Target Type classifies the entity assayed (single protein, protein complex, organism/virus, cell line, or other) and Assay Type follows the ChEMBL convention: B (binding/target inhibition), F (functional/cell-based), A (ADME), P (physicochemical), T (toxicity). Both were taken from the ChEMBL target annotation where a ChEMBL identifier was available (including cross-referenced PubChem entries); BindingDB records were assigned Assay Type = B and a Target Type from its protein-chain count (one → single protein, more than one → protein complex). Otherwise the fields were inferred from the assay name/description and flagged with an asterisk to mark their provenance. Cytotoxicity and assay-interference were differentiated from selective antiviral activity. Selectivity indices (SI = CC50/EC50, same cell line) were computed where paired data existed (133 compounds), and all compounds were flagged for likely cytotoxicity (Cytotoxic flag; 91 compounds, by ATC-L01 class, cytotoxic mechanism [topoisomerase/tubulin], or SI < 10) and for pan-assay interference (Baell & Holloway, 2010) (PAINS; 1,529 compounds, 11.6%). Remaining compounds were grouped by identical SMILES and target name and flagged in the Assay Variability Note column when the maximum-to-minimum activity ratio against the same target was 10 or higher. Extraction accuracy was verified manually for 200 randomly selected compounds, covering 509 molecule–assay records. The sample size was fixed in advance using Cochran's formula (Cochran, 1977, Ch. 4) with a finite population correction (Z = 1.96, p = 0.5, e = 0.05, N = 13,194), which required 374 records (Supplementary Method 3).

### 2.4. Cheminformatic analysis

For further chemoinformatic analysis, compounds were standardized. The standardization is described in Supplementary Method 4. Nucleoside analogues were identified (Supplementary Method 5 and Supplementary Note 8). Eight physicochemical descriptors and QED and HADES drug-likeness scores were computed, with HADES training compounds excluded (Supplementary Method 6). Ring systems and Bemis-Murcko scaffolds (Supplementary Method 7), generalized scaffolds obtained by DBSCAN clustering of cyclic skeletons (Ester et al., 1996) (Supplementary Method 8), scaffold-based substructure searches with positional activity prevalence (Supplementary Method 9), structural clusters derived from ECFP4 fingerprints projected by UMAP (McInnes et al., 2018) and clustered with HDBSCAN (Supplementary Method 10), and functional and alerting group analysis (Supplementary Method 11) were performed as described.

## 3. Results

### 3.1. Dataset preparation with LLM

Antiviral compounds were extracted for a total of 23 viruses, as detailed in Section 2 (Fig. 1a). From BindingDB (release 202601, downloaded January 2026), 1,390,080 compounds were initially retrieved; however, relevant data were identified for only nine viruses, CHIKV, DENV, EBOV, HTNV, IV, MERS-CoV, SARS-CoV, NiV, and ZIKV, yielding a total of 27,364 extracted compounds. From ChEMBL (release 36, downloaded January 2026), 47,886 bioassay records under the "Viruses" category were retrieved. As this category spans all viruses in ChEMBL, only records for our viruses of interest were retained. SMILES strings and activity values were extracted for each virus; no data were available for CCHFV and SUDV. This yielded 12,034 bioassays and 57,560 compounds. From PubChem (downloaded December 2024), 41,273 bioassay records were retrieved using targeted search queries. After extraction of virus names from bioassays with DeepSeek LLM (DeepSeek-AI, 2024; Section 2.2), 16,364 bioassays and 3,601,919 compounds with associated SMILES strings and activity data were obtained. Because PubChem records are not manually curated, mismatches and incomplete metadata frequently occur. Data-related problems affecting the PubChem extraction for viruses, along with the corresponding numerical values, are reported in Supplementary Table 1.

To validate this extraction of the selected viruses, DeepSeek's accuracy was benchmarked against expert-curated ChEMBL annotations. Because every benchmarked assay has one curated reference label and receives one extracted label, accuracy is the appropriate metric.Extraction accuracy was 96.15-100% across 20 virus-specific ChEMBL assay sets, with residual errors confined to unresolved serotypes and to viral vectors being reported in place of the target. Across 1,000 PubChem assays, each processed five times, mean per-field consistency was 0.96–0.97. Full benchmarking and reproducibility results, including per-virus accuracies, the error categories and per-field consistency scores, are provided in Supplementary Note 1 and Supplementary Table 2.

## 3.2. Final dataset construction and evaluation

Following target name identification via DeepSeek and merging of the datasets from all three databases, a combined dataset of 3,686,843 compounds was obtained. After deduplication, 3,440,391 compounds remained. Following conversion of activity values to nanomolar units and removal of records that could not be converted, 1,439,886 molecules remained. Finally, after excluding compounds with activity values greater than 10,000 nM, the dataset comprised 30,714 compounds. The number of extracted compounds and bioassays for each virus across all three databases is summarized in Supplementary Table 3. Following grouping of data items, as described in Section 2.3, the BSA-DB was obtained, comprising 13,194 compounds . HTNV, the one virus from the 23 selected with no compound meeting this criterion, is absent from the final database (Fig. 1a). These compounds exhibited antiviral activity, defined as ≤10000 nM, against two or more viruses. Only 79 compounds were identified as having at least one *Yes* entry in the *Assay Variability Note* column, i.e., a ratio of ≥10 between the maximum and minimum activity values reported against the same target, indicating high intra-group variability.

To verify extraction accuracy (SMILES, Standard Type, Relation, Value, and Unit), 200 randomly selected compounds from the final database, covering 509 bioassays, were manually validated (Supplementary Table 4). The sample size was determined in advance by Cochran's formula with a finite population correction, which required 374 records (Supplementary Method 3). Only 8 of the 509 bioassays (1.57%) contained mismatches: in AID720532, Perhexiline maleate was extracted without the maleate moiety; in AID1117302, the SMILES extracted for Etoposide corresponded to CHEMBL3348857, a stereoisomer. Compound identity and standard activity values were likewise misextracted in AID720532 and AID720533. As these involve minor structural variants (salt forms, stereoisomers) rather than totally incorrect compounds, and are confined to a few AIDs, they do not substantially compromise the reliability and accuracy of the final database.

## 3.3. Landscape of broad-spectrum antiviral activity in BSA-DB

The composition of BSA-DB such as compound counts, potency ranges, and assay-type distribution is summarized per virus in **Table 1**.

| **Virus Name** | **Unique Compounds** | **Measurements** | **Functional (%)** | **Binding (%)** | **Other (%)** | **<10 nM (%)** | **10–100 nM (%)** | **100–1,000 nM (%)** | **1,000–10,000 nM (%)** |
|---|---|---|---|---|---|---|---|---|---|
| MARV | 10,326 | 10,383 | 99.6 | 0.4 | 0.0 | 0.5 | 1.5 | 16.1 | 82.0 |
| LASV | 9,432 | 9,557 | 99.8 | 0.2 | 0.0 | 1.1 | 1.5 | 15.2 | 82.2 |
| IV | 3,861 | 4,083 | 9.7 | 90.2 | 0.1 | 0.7 | 0.5 | 7.8 | 91.0 |
| SARS-CoV | 2,702 | 4,287 | 21.4 | 77.1 | 1.4 | 6.7 | 9.8 | 21.9 | 61.5 |
| EBOV | 436 | 607 | 73.3 | 25.5 | 1.2 | 3.0 | 4.1 | 24.7 | 68.2 |
| ZIKV | 426 | 615 | 36.9 | 63.1 | 0.0 | 3.3 | 7.0 | 24.1 | 65.7 |
| MERS-CoV | 409 | 576 | 28.1 | 71.7 | 0.2 | 18.4 | 20.5 | 36.5 | 24.7 |
| JUNV | 88 | 107 | 79.4 | 20.6 | 0.0 | 59.8 | 19.6 | 7.5 | 13.1 |
| DENV | 77 | 124 | 0.0 | 100.0 | 0.0 | 4.0 | 5.6 | 26.6 | 63.7 |
| CHIKV | 57 | 149 | 83.2 | 14.8 | 2.0 | 4.7 | 7.4 | 37.6 | 50.3 |
| WNV | 44 | 49 | 0.0 | 100.0 | 0.0 | 0.0 | 2.0 | 28.6 | 69.4 |
| YFV | 36 | 40 | 100.0 | 0.0 | 0.0 | 0.0 | 7.5 | 42.5 | 50.0 |
| RVFV | 20 | 41 | 100.0 | 0.0 | 0.0 | 0.0 | 29.3 | 36.6 | 34.1 |
| MPXV | 12 | 20 | 90.0 | 0.0 | 10.0 | 0.0 | 50.0 | 20.0 | 30.0 |
| TBEV | 11 | 20 | 95.0 | 0.0 | 5.0 | 0.0 | 30.0 | 10.0 | 60.0 |
| VEEV | 11 | 12 | 100.0 | 0.0 | 0.0 | 0.0 | 0.0 | 8.3 | 91.7 |
| VARV | 10 | 24 | 100.0 | 0.0 | 0.0 | 0.0 | 58.3 | 25.0 | 16.7 |
| NiV | 4 | 11 | 100.0 | 0.0 | 0.0 | 0.0 | 63.6 | 9.1 | 27.3 |
| VACV | 3 | 3 | 0.0 | 100.0 | 0.0 | 0.0 | 0.0 | 0.0 | 100.0 |
| SFTSV | 2 | 2 | 100.0 | 0.0 | 0.0 | 0.0 | 0.0 | 100.0 | 0.0 |
| CCHFV | 1 | 3 | 0.0 | 100.0 | 0.0 | 0.0 | 0.0 | 100.0 | 0.0 |
| SUDV | 1 | 1 | 100.0 | 0.0 | 0.0 | 0.0 | 0.0 | 0.0 | 100.0 |

**Table 1. Summary of BSA-DB composition by virus.** For each virus, the table reports the number of unique compounds, the number of compound–assay measurements, the assay-type distribution, and the distribution of reported potencies. *Unique Compounds*, distinct compounds with reported activity (≤ 10,000 nM) against the virus; *Measurements*, the number of compound-assay records contributing those values. *Functional*, *Binding*, and *Other* give the percentage of measurements by assay type following the ChEMBL convention (Functional = cell-based/phenotypic; Binding = target binding or target inhibition; Other = ADME, physicochemical, or toxicity assays). The four potency columns give the percentage of measurements falling in each concentration range (< 10, 10–100, 100–1,000, and 1,000–10,000 nM). Because BSA-DB is a broad-spectrum resource, compounds active against more than one virus are counted under each virus they inhibit; the *Unique Compounds* column therefore does not sum to the 13,194 compounds in the database. Overall, BSA-DB comprises 13,194 compounds and 30,714 measurements (73% functional, 27% binding; potency < 10 / 10–100 / 100–1,000 / 1,000–10,000 nM = 2.2 / 3.3 / 16.5 / 78.0%). Percentages are not meaningful for viruses with fewer than five measurements (VACV, SFTSV, CCHFV, SUDV).

To characterize the broad-spectrum compounds in the BSA-DB, we first identified the most common virus family combinations and then the most frequently represented compounds.

Eight of the top 10 combinations involved viruses from different families. The most represented was LASVMARV (*Arenaviridae* and *Filoviridae*), with 6,924 compounds showing inhibitory activity against both viruses, followed by IV-MARV (*Orthomyxoviridae* and *Filoviridae*) with 1,463 compounds, IV-LASV with 848, IV-SARS-CoV (*Coronaviridae*) with 623, and MARV-SARS-CoV with 578. The two remaining combinations involved viruses from the same family: MERS-CoV-SARS-CoV with 366 compounds and EBOV-MARV with 148. Several three-virus combinations were also identified, indicating that many compounds inhibit viruses spanning three families: IV-LASV-MARV contained 566 compounds and LASV-MARV-SARS-CoV contained 261.

We then identified the compounds with the highest virus coverage (Fig. 1b). Remdesivir, developed originally against hepatitis C and later approved for COVID-19 (Beigel et al., 2020), had the widest spectrum, active against nine viruses: EBOV, JUNV, LASV, MERS-CoV, NiV, SARS-CoV, TBEV, YFV and ZIKV. Mycophenolic acid and niclosamide ranked second and third, each active against seven viruses, acting largely through host processes: purine nucleotide metabolism (Allison and Eugui, 1996) and endosomal acidification (Jurgeit et al., 2012), respectively. Doxorubicin, OSU-03012, CHEMBL5273609 and lycorine each covered six viruses; of these, only CHEMBL5273609, an isomer of galidesivir (Sparrow et al., 2023), was a nucleoside analogue, as was remdesivir. Interestingly, we identified only 291 nucleoside analogues (2.2% of the BSA-DB compounds) (Supplementary Table 5). Most were active against LASV (n = 211), MARV (n = 210), IV (n = 63), SARS-CoV (n = 55) and EBOV (n = 25). Breadth beyond two viruses was rare: 20 compounds were active against three, 10 against four, and one each against five, seven and nine. The most potent molecule was CHEMBL5205032 ([BSADB_1888]), with an EC50 of 1 nM against SARS-CoV-2. It was followed by SCHEMBL20834450 ([BSADB_990]), which reached 1.4 nM (EC50) against DENV, and CHEMBL3797928 ([BSADB_12644]), which showed 2 nM (EC50) against influenza A virus (H1N1). Median potency was 2.2 µM. Seventeen analogues are approved antiviral or antineoplastic drugs, including gemcitabine and cytarabine (Fig. 1c) (Supplementary Table 6).

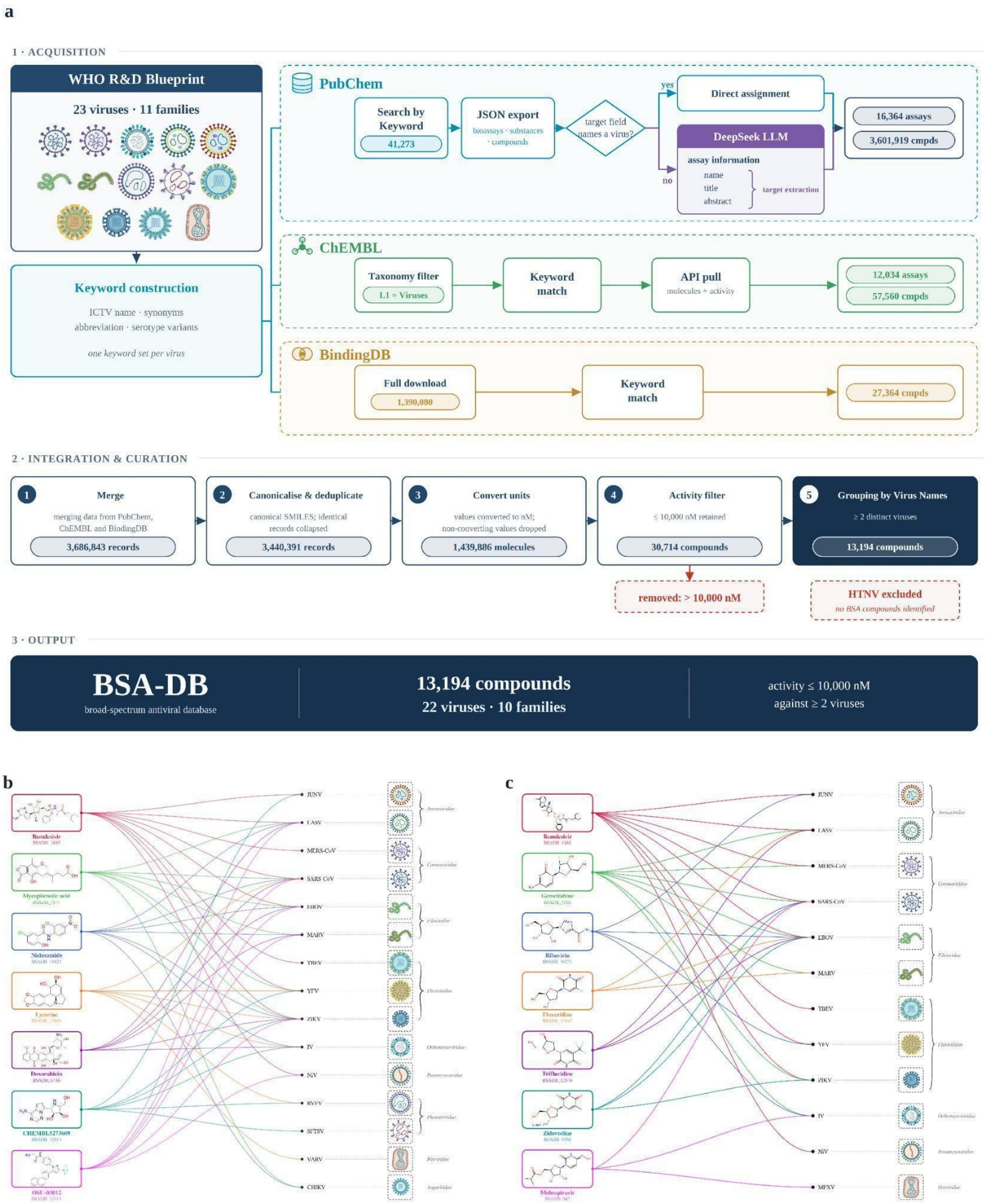

**Fig. 1: BSA-DB creation workflow and broad-spectrum antiviral chemotypes and their reported virus coverage.**

**a,** Construction of BSA-DB. Each of the 23 viruses prioritized in the WHO R&D Blueprint was assigned a keyword set (ICTV species name, synonyms, standard abbreviation and serotype variants) used to query PubChem, ChEMBL and BindingDB. PubChem assays whose target field explicitly named a virus were assigned directly; the remainder were resolved by LLM extraction of the target from the assay name, title and abstract. Merged records were canonicalized and deduplicated, converted to nM, filtered at ≤10,000 nM, and grouped by virus, retaining 13,194 compounds with reported activity against ≥2 of 22 viruses spanning 10 families. HTNV yielded no compounds meeting these criteria and is absent from the final database. **b,** Bipartite networks linking compounds (left) to viruses with reported activity (centre), grouped by family (right). Curve colour identifies the compound; structures are labelled with name and BSADB ID. The seven most broad-spectrum antiviral molecules in BSA-DB, covering 15 viruses across nine families. **c**, The seven most broad-spectrum approved nucleoside analogues, covering 12 viruses across seven families, ranked by virus count then by the most potent reported value (AC50, EC50, EC90, IC50 or potency). Remdesivir appears in both panels in the same colour. Virion schematics are illustrative, not to scale, and the poxvirus icon is shared between VARV (**b**) and MPXV (**c**). Structures were rendered with RDKit (https://www.rdkit.org).

## 3.4. Physicochemical and structural characterization

To evaluate the physicochemical profile of BSA-DB relative to clinically validated therapies, we compared the dataset with a benchmark set of approved antiviral drugs (Fig. 2a). BSA-DB compounds were smaller and more lipophilic, whereas approved antivirals were more polar, richer in hydrogen-bond donors and acceptors, richer in sp3-hybridized carbons and more flexible. Ring counts were comparable. Full numerical details are provided in Supplementary Table 7.

Scaffold and ring system analysis was performed to characterize the structural diversity of BSA-DB. The ten most frequent ring systems, excluding benzene, are shown in Fig. 2b. Pyridine was the most prevalent (8.87% of compounds), followed by furan (6.74%), thiophene (6.40%), piperazine (5.97%), cyclohexane (5.54%) and morpholine (4.06%), along with several fused heterocyclic systems. Pairwise similarity based on maximum common substructure (MCS) was then calculated for the top 100 Murcko scaffolds (Bemis and Murcko, 1996) to assess structural redundancy (Fig. 2c). The matrix revealed distinct scaffold clusters, but many scaffold pairs scored low, indicating that the scaffold space remains broadly diverse. The three most abundant Murcko scaffolds (Fig. 2d) were a benzamide-diphenyl framework (n = 59, 0.45%), a diphenyl ether scaffold (n = 35, 0.27%) and a diphenyl-imidazo[1,2-a]pyridine scaffold (n = 32, 0.24%). That even these account for well under 1% indicates that no single framework dominates the BSA-DB.
Then, we extracted promising, generalized scaffolds based on an initial potency threshold of <10 nM against all target viruses, subsequently refining them through an MCS similarity filter (>0.7). They were considered promising because their recurrence in multiple potent derivatives demonstrates that the scaffold consistently supports sub-10 nM inhibition, establishing a robust foundation for analogue optimization. After extraction, their chemical space was analyzed (Supplementary Table 8). Each of the nine retained scaffolds was analysed for its positional activity prevalence (Supplementary Figures 1–8), which measures whether substituents at a given position derive predominantly from active or inactive compounds. Compounds were classified as active against a given target at a threshold of 10 µM (10,000 nM). This normalized metric ranges from 0 to 1, where a value of 0 indicates exclusive presence in inactive compounds, 1 indicates exclusive presence in active compounds, and 0.5 represents equal representation across both classes. The scaffolds differed in the amount of positional information they could support. Scaffold S1 was recovered from 15 of 23 targets and was the only motif populating both activity classes across most targets (Fig. 2e). S1 consists of a two-ring core with a two-atom linker. JUNV, YFV, and RVFV were excluded from all positional deviation calculations, as their prevalence values are only 0.00 or 1.00 and reflect that the activity of a compound does not depend on substituent position; the same exclusion applies throughout the analyses below. Substitutions at the para positions (atoms 5 and 11) of the S1 scaffold were nearly equally found in active and inactive compounds across targets, with mean deviations from the neutral value of 0.5 of only 0.075 and 0.052, respectively. In contrast, at the ortho position of the aromatic ring (atom 3) and the meta position of the second ring (atom 10), the proportion of active versus inactive compounds varied more significantly across targets, as evidenced by larger mean deviations from the neutral 0.5 value: 0.171 and 0.191, respectively. There are also target-based patterns. Flaviviruses exhibited reciprocal profiles: some positions that are more frequently found in active compounds for DENV are in inactive compounds for WNV and ZIKV, and vice versa, where positions active for WNV and ZIKV are inactive for DENV. DENV profiles correlated inversely with those of WNV ($r = -0.87$, $n = 13$, 95% CI [−0.96, −0.61], $q = 0.0008$) and ZIKV ($r = -0.65$, $n = 12$, 95% CI [−0.89, −0.12], $q = 0.022$), whereas WNV and ZIKV correlated positively ($r = +0.88$, $n = 12$, 95% CI [+0.62, +0.97], $q = 0.0003$). All three associations remained significant after correction for multiple comparisons (Supplementary Method 13, Supplementary Table 8). EBOV showed that some sites are more common in active compounds and others in inactive compounds. Profiles for LASV, MARV, and SARS-CoV for all sites were nearly flat, with mean deviations from 0.5 of 0.035, 0.033, and 0.016, respectively, indicating no activity difference regarding substituents.

LASV exhibited position-dependent activity prevalence on scaffold S4. Nevertheless, on scaffold S6, substituents at most positions are mostly found in inactive compounds. Positions of substituents on scaffolds S3 and S7–S9 are found only in active or only in inactive compounds, reflecting that they do not have any substitution preferences (Supplementary Figures 1–8).

Virtual screening of PubChem across the nine scaffolds identified over 1.55 million total matches, though analogue availability was heavily skewed toward just three primary structures (Supplementary Note 2 and Supplementary Table 9).

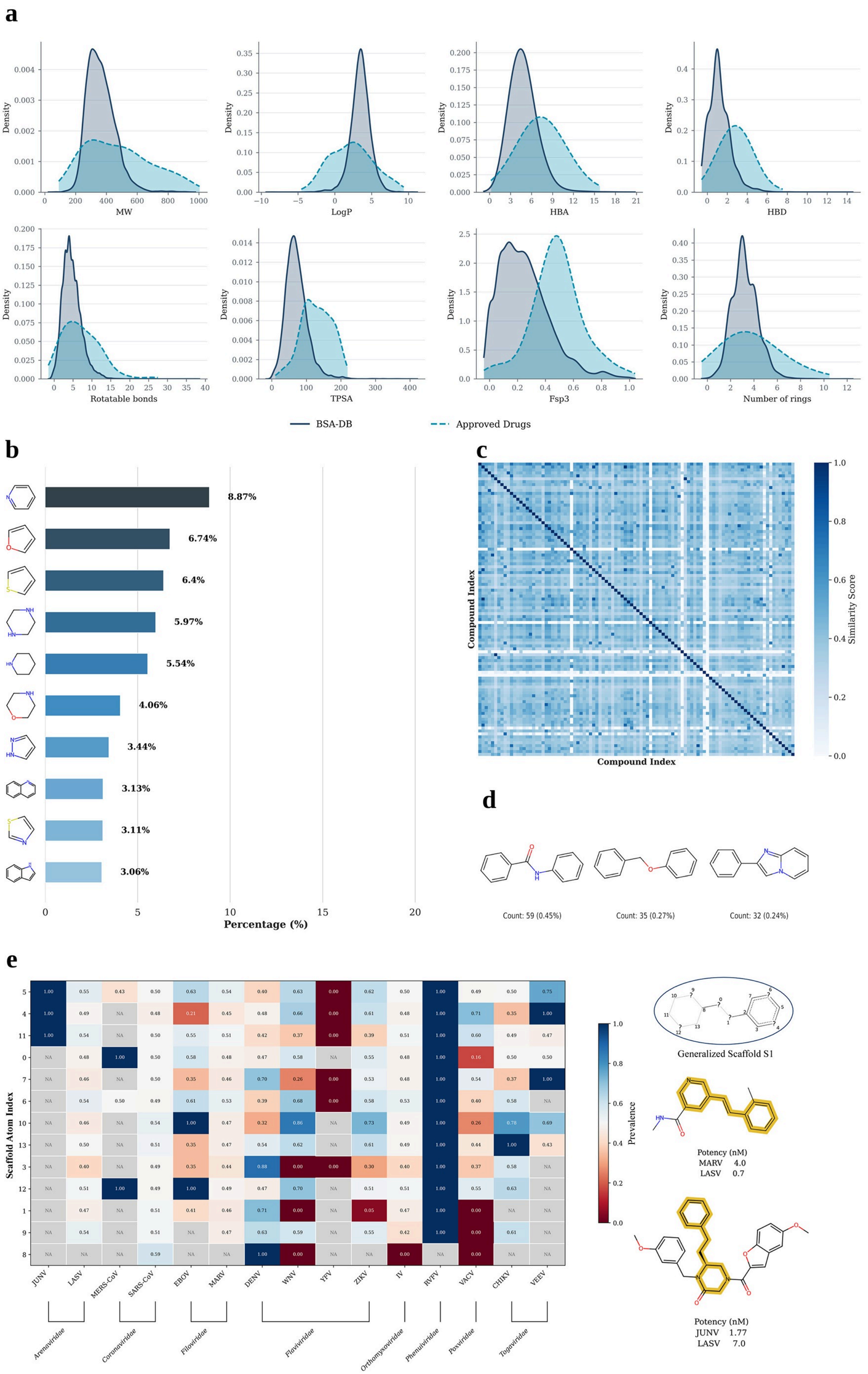
a
Density
MW
LogP
HBA
HBD
Rotatable bonds
TPSA
Fsp3
Number of rings
BSA-DB
Approved Drugs
b
8.87%
6.74%
6.4%
5.97%
5.54%
4.06%
3.44%
3.13%
3.11%
3.06%
Percentage (%)
c
Compound Index
Compound Index
Similarity Score
d
Count: 59 (0.45%)
Count: 35 (0.27%)
Count: 32 (0.24%)
e
Scaffold Atom Index
JUNV
LASV
MERS-CoV
SARS-CoV
EBOV
MARV
DENV
WNV
YFV
ZIKV
IV
RVFV
VACV
CHIKV
VEEV
Arenaviridae
Coronaviridae
Filoviridae
Flaviviridae
Orthomyxoviridae
Phenuiviridae
Poxviridae
Togaviridae
Prevalence
Generalized Scaffold S1
Potency (nM)
MARV 4.0
LASV 0.7
Potency (nM)
JUNV 1.77
LASV 7.0

**Fig. 2: Physicochemical, scaffold, and positional activity landscape of BSA-DB.**

**a**, Distributions of eight physicochemical properties for BSA-DB compounds (grey) and approved antivirals (blue). **b**, Frequency of the ten most common non-benzene ring systems. **c**, MCS pairwise similarity matrix for the top 100 molecular scaffolds. **d**, The three most abundant Murcko scaffolds. **e**, Positional activity prevalence for each atom position of the generalized scaffold S1 (columns) across viral targets (rows). Prevalence is bounded in the interval [0, 1]: 1 (blue) marks positions with attachments only in active compounds, 0 (red) only in inactives, and 0.5 (white) equal attachments in both. Grey (NA) cells indicate positions that do not have any attachments. Columns are grouped by viral families. Right: The indexed scaffold S1 and two representative compounds with the shared core, alongside their potencies (nM) against the indicated viruses.

## 3.5. Structural clustering

Structural clustering was performed to identify groups of related compounds and characterize chemical diversity within BSA-DB. We applied HDBSCAN (McInnes et al., 2017) to the 13,194 compounds, identifying 497 clusters under the parameters selected by grid search (min_cluster_size = 5, min_samples = 5; DBCV = 0.304) (Moulavi et al., 2014). Cluster −1, comprising 4,523 noise points (34.3%), was excluded, as these compounds were structurally unique and did not fall within any dense chemical neighbourhood. Compounds targeting MARV, LASV, IV and SARS-CoV appeared across almost all clusters, consistent with these being the four best-represented viruses (Fig. 3a). By contrast, compounds inhibiting NiV, SUDV and CCHFV were each confined to a single cluster, reflecting their limited representation.

Cluster sizes varied considerably, the largest being cluster 356 with 136 compounds. Pairwise MCS similarity across this cluster was mostly low to moderate, but a few compact off-diagonal blocks of high similarity were evident, indicating discrete groups of near-identical analogues within an otherwise heterogeneous cluster (Fig. 3b). Representative pairs included two ester-substituted, fused bicyclic pyrrole analogues differing only in their amide aryl substituent (BSADB_1204/BSADB_1205; MCS similarity 0.985), and two pairs of methylbenzoyl-substituted triazolopyridine-thione derivatives differing only in the position of ring methylation (BSADB_8785/BSADB_8786 and BSADB_8785/BSADB_9122; 0.974 for both) (Fig. 3c). Consistently, scaffold enumeration of cluster 356 identified an N-benzylindole core as the most recurrent scaffold (n = 6), followed by a phenylthiophene-carboxamide scaffold bearing a morpholinopropyl substituent and a phenyl-substituted triazolopyridine-thione scaffold, each with three compounds (Fig. 3d). Cluster 356 is therefore not chemically uniform despite its size, but comprises several small series of close analogues built around a few recurring scaffolds. Despite this one large cluster, BSA-DB is dominated by considerably smaller clusters, containing fewer than 20 compounds, and 25 clusters held the minimum of 5 compounds (clusters 42, 154, 157, 181,

206, 208, 211, 227, 228, 271, 282, 306, 321, 365, 368, 396, 403, 413, 416, 433, 456, 461, 462, 466, 478) (Fig. 3e).

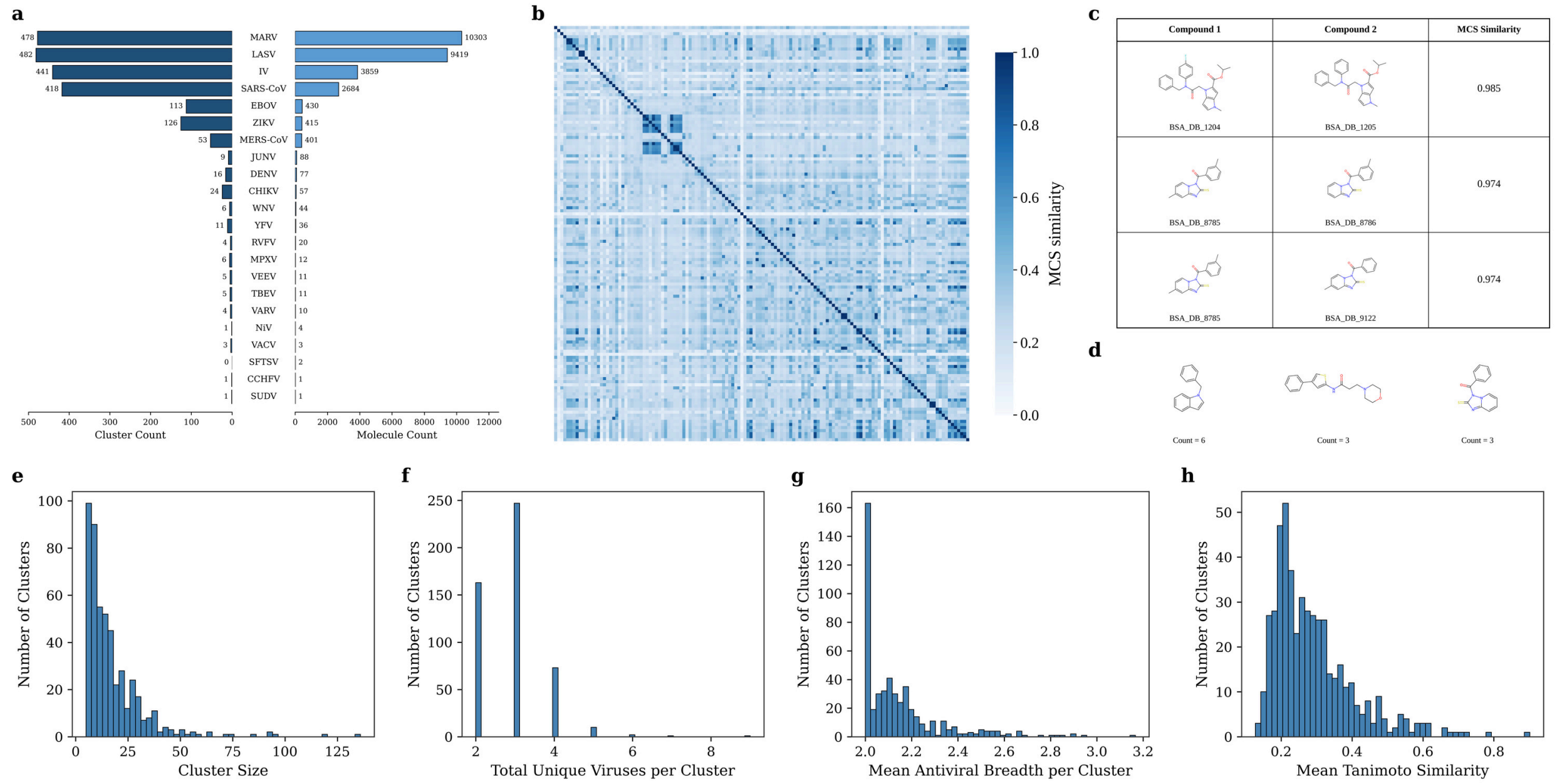

**Fig. 3: Chemical landscape of the broad-spectrum antiviral dataset.**

**a,** Chemical clusters (left, navy) and total molecules (right, blue) per virus, ranked highest to lowest. SFTSV's molecules fall within the noise cluster. **b**, Pairwise MCS similarity matrix between the biggest cluster (cluster 356) representatives. Off-diagonal blocks mark clusters sharing a common core; the pale background indicates most clusters are structurally distinct. **c,** The three most similar cross-cluster (cluster 356) compound pairs (MCS > 0.97), with structures, BSA-DB IDs, and similarity values. Each pair differs by a single substituent. **d**, The three most recurrent molecular scaffolds, with the number of clusters (cluster 356) containing each. **e**, Cluster size distribution across 497 clusters. **f,** Number of unique viruses per cluster. **g**, Mean antiviral breadth per cluster (average viruses targeted per compound). **h,** Mean pairwise Tanimoto similarity within clusters (ECFP4, radius = 2, 2048 bits), indicating structural homogeneity. All analyses excluded noise points (cluster −1, n = 4,523). Compound counts in these figures refer to the standardized analysis set and therefore differ slightly from the full curated dataset.

To identify clusters with the highest broad-spectrum potential, clusters were ranked by the total number of unique target viruses. As shown in Fig. 3f, most clusters covered up to 6 viruses. However, two clusters stood out: cluster 14 (n = 15 compounds) and cluster 5 (n = 18 compounds), each covering 14 unique viruses. Both clusters shared common viruses such as CHIKV, EBOV, IV, JUNV, LASV, MARV, MERS-CoV, RVFV, SARS-CoV, YFV, and ZIKV. Cluster 5 additionally contains DENV, NiV, and TBEV, which were absent from cluster 14. Cluster 14, in turn, contains MPXV, SUDV, and VARV, representing poxviruses and filoviruses not present in cluster 5. Both clusters are dominated by nucleoside analogues, a remdesivir-like series in cluster 5 (including prodrugs, active metabolites and stereoisomers of one another) and a carbocyclic nucleoside series in cluster 14 containing several clinically explored antivirals (Supplementary Note 3).

Clusters were also ranked by mean antiviral breadth, defined as the average number of distinct viruses targeted per compound within the cluster. The highest mean antiviral breadth was observed in cluster 54 (n = 6 compounds; mean antiviral breadth = 3.17), which is composed of Vinca alkaloids, a class of naturally derived compounds originally isolated from *Catharanthus roseus* and used in oncology  (van der Heijden et al., 2004). These compounds showed activity against multiple viruses (ZIKV, LASV, EBOV, MARV, and SARS-CoV). Because Vinca alkaloids disrupt microtubule dynamics and are generally cytotoxic to host cells, their broad-spectrum profile should be interpreted with care. Importantly, however, the reported antiviral activities were measured at non-toxic (sub-cytotoxic) concentrations, so this multi-virus activity does not simply reflect interference with host-cell viability (Supplementary Note 3). The second-highest mean antiviral breadth cluster was cluster 5 (mean antiviral breadth = 2.94), described in Supplementary Note 3. The third-highest mean antiviral breadth was observed in cluster 2 (n = 9; mean breadth = 2.89), which consisted exclusively of camptothecin derivatives. Camptothecins are alkaloids originally isolated from Camptotheca acuminata and are known inhibitors of DNA topoisomerase I, with several derivatives approved for cancer treatment (Pommier, 2006). Cluster 2 contained irinotecan (BSADB_3657), active against four viruses, and several camptothecin derivatives (Supplementary Note 3). As with the Vinca alkaloids, this cluster is structurally homogeneous and consists of topoisomerase I inhibitors with well-established cytotoxicity. All members of clusters 54 and 2 are accordingly flagged in the database as cytotoxic (Cytotoxic field; by ATC-L01 anticancer classification, cytotoxic mechanism of action [tubulin or topoisomerase], and/or low selectivity index, SI < 10). Cytotoxicity does not by itself negate the broad-spectrum antiviral value of these compounds, as their reported activities were obtained at non-toxic concentrations; rather, it indicates that such compounds should be prioritized cautiously and evaluated carefully for their therapeutic window in animal models before further development.

To assess whether the compounds within each cluster were structurally homogeneous or heterogeneous, mean pairwise Tanimoto similarity across cluster members was calculated for each cluster (Fig. 3h). For most clusters, the mean Tanimoto similarity was below 0.3, indicating high structural diversity among the compounds within each cluster. The three clusters with the highest mean pairwise Tanimoto similarity across cluster members were cluster 29 (n = 7, mean Tanimoto = 0.901), cluster 38 (n = 7, mean Tanimoto = 0.787), and cluster 41 (n = 12, mean Tanimoto = 0.724), indicating that these clusters are composed of structurally highly homogeneous compounds.

The table of all clusters (excluded noise) with their cluster size, total unique viruses per cluster, mean antiviral breadth per cluster, and mean Tanimoto similarity is provided in Supplementary Table 10.

## 3.6. Drug-likeness analysis

Although physicochemical properties are informative, they are insufficient to fully assess the drug-likeness of compounds. We therefore calculated two drug-likeness metrics for all BSA-DB compounds: quantitative estimate of drug-likeness (QED) (Bickerton et al., 2012) and the machine-learning-based holistic AI-based drug-likeness estimation score (HADES) (Petrosyan et al., 2026). Compounds used in HADES training were excluded from the score distributions, which therefore comprise 13,026 BSA-DB compounds and 30 approved antiviral drugs (Fig. 4a). Notably, the HADES training set contained 49 additional approved antiviral drugs (0.8% of the training set). The mean QED and HADES scores of approved antiviral drugs were 0.408 and 0.677, respectively, while the corresponding mean values for BSA-DB compounds were 0.599 and 0.140.

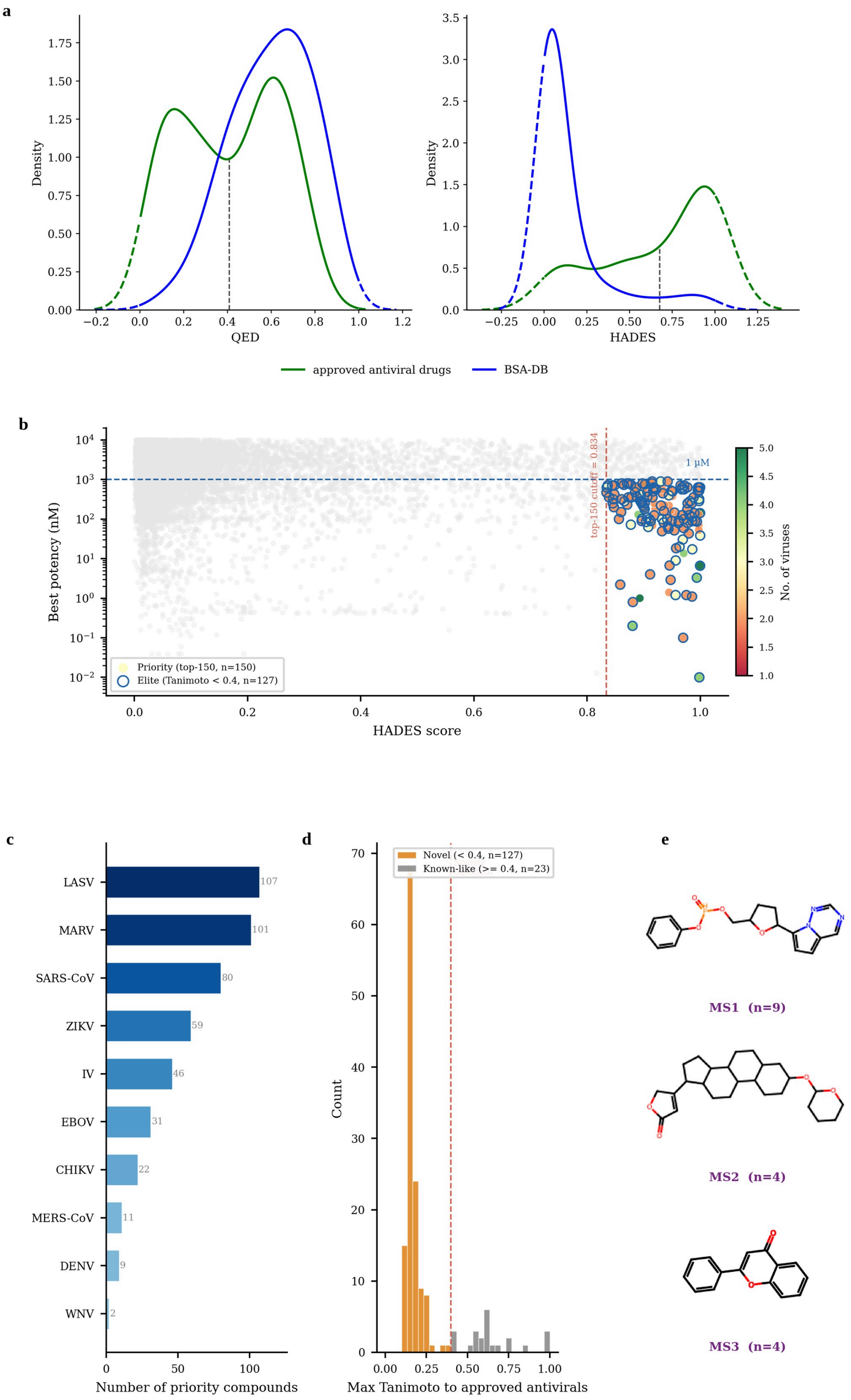
a
Density
QED
HADES
approved antiviral drugs
BSA-DB
b
Best potency (nM)
HADES score
top-150 cutoff = 0.834
1 µM
No. of viruses
Priority (top-150, n=150)
Elite (Tanimoto < 0.4, n=127)
c
LASV
MARV
SARS-CoV
ZIKV
IV
EBOV
CHIKV
MERS-CoV
DENV
WNV
107
101
80
59
46
31
22
11
9
2
Number of priority compounds
d
Novel (< 0.4, n=127)
Known-like (>= 0.4, n=23)
Count
Max Tanimoto to approved antivirals
e
MS1 (n=9)
MS2 (n=4)
MS3 (n=4)

**Fig. 4: Identification and multi-level characterization of novel broad-spectrum antiviral candidates and their privileged scaffolds in BSA-DB.**

**a,** KDE distributions of QED (left) and HADES (right) scores for approved antiviral drugs (green, $n$ = 30) and BSA-DB compounds (blue, $n$ = 13,026); dashed grey lines indicate the approved-drug means (QED = 0.408; HADES = 0.677). **b,** HADES score versus best recorded potency for eligible BSA-DB compounds. The 150 highest-scoring compounds with potency <1 µM constitute the priority subset; point colour indicates the number of viruses, and blue outlines denote elite compounds with maximum Tanimoto similarity <0.4 to approved antivirals ($n$ = 127). Dashed lines indicate the potency and HADES cutoffs. **c,** Virus representation among priority compounds, showing the ten most represented viruses. **d,** Maximum Tanimoto similarity of priority compounds to approved antivirals; orange, novel ($n$ = 127); grey, known-like ($n$ = 23). The dashed line marks the 0.4 threshold. **e,** Three most populated Murcko scaffold classes among priority compounds: MS1 ($n$ = 9), MS2 ($n$ = 4), and MS3 ($n$ = 4).

To decide which of the two metrics better reflects clinical viability, both were benchmarked on the 415 BSA-DB compounds that could be cross-referenced to ChEMBL entries with a recorded clinical development phase. After removal of HADES training compounds (n = 137), 278 held-out compounds remained, of which 103 (37.1%) were approved drugs. On this held-out set, HADES separated approved from non-approved compounds with Receiver Operating Characteristic - Area Under the Curve (ROC-AUC) score equal to 0.70, whereas QED performed no better than chance (ROC-AUC 0.47). HADES was therefore used for subsequent prioritisation.

To identify promising candidates beyond existing clinical pipelines, these 415 clinically annotated compounds were excluded together with a further 31 HADES training compounds, leaving 12,748. Rather than applying a score threshold derived from approved antivirals, whose HADES scores are inflated by training-set inclusion, a fixed screening budget was applied: the 150 highest-scoring potent compounds were defined as the priority subset (Fig. 4b). Because most approved antivirals were present in the HADES training set, their mean score fell from 0.876 to 0.677 on their removal, making it an unreliable anchor; the resulting rank-based cutoff (HADES = 0.834) remains well above this de-contaminated mean, and the novel fraction and MS1–MS3 membership are unchanged across N = 50-200 (Supplementary Table 11). Collectively, these were active across 15 viruses, with LASV, MARV and SARS-CoV most frequently represented (107, 101 and 80 compounds, respectively; Fig. 4c).

Structural novelty of the priority compounds was assessed as maximum Tanimoto similarity to approved antiviral drugs. The majority exhibited maximum similarity below 0.2, with the distribution strongly concentrated near 0.1–0.15, indicating that these compounds are not structural analogues of known antivirals (Fig. 4d). Applying a novelty threshold of Tanimoto < 0.4 (Stokes et al., 2020), 127 compounds (84.7%) were designated as elite candidates — simultaneously drug-like, potent, broad-spectrum, and structurally distinct from any approved antiviral drug. Murcko scaffold (MS) analysis of the priority subset revealed three chemically distinct scaffold classes (Fig. 4e), designated MS1–MS3, with MS1 the most frequent (n = 9), followed by MS2 and MS3 (n = 4 each). These 127 compounds are computationally prioritised candidates requiring experimental validation rather than established antiviral leads; HADES discrimination on held-out data is moderate (ROC-AUC 0.70), and novelty is assessed only against approved antivirals, so similarity to investigational or preclinical chemotypes is not excluded.
A detailed evaluation of the breadth of antiviral coverage, viral family mapping, and experimental assay validation for these privileged scaffolds is provided in Supplementary Note 4.

## 3.7. Antiviral potency distributions across viruses

The distribution of curated activity values across the seven best-represented viruses is shown in Fig. 5a. Where a compound had multiple measurements against the same virus, the geometric mean was used; single measurements were taken as reported. For six of seven viruses, potency values concentrated above 100 nM, placing this cut-off in the sparse left tail rather than the densest region. Compounds ≤ 100 nM were classified as highly potent and those > 100 nM as potent; the highly potent proportion varied over 90-fold across viruses. MERS-CoV showed the highest potent fraction, with 186 of 409 compounds (45.5%) below 100 nM, and its distribution was distinctly bimodal, with a well-resolved sub-100 nM mode absent from all other viruses. Because such a highly potent fraction could in principle reflect a single screening campaign, we examined the assay provenance of the MERS-CoV compounds on either side of the threshold. Compounds ≤100 nM were drawn from 140 unique assays (195 assay–compound records in total), with the most heavily represented assay (AID 1974283) contributing only 18 compounds (9.68%); compounds >100 nM were drawn from 112 unique assays (278 records), with the two largest assays (ChEMBL5168247 and AID 1974283) each contributing 20 compounds (8.97%). The comparable and high assay diversity on both sides of the threshold indicates that the elevated potent fraction for MERS-CoV is not an artefact of a single high-throughput campaign, but is supported by consistent sub-100 nM activity across many independent assays, although distinct assay identifiers may share a publication or depositor.  This conclusion is further reinforced at the structural level: the pairwise MCS similarity distribution among MERS-CoV high potent compounds (mean = 0.33, median = 0.2) was strongly right-skewed, with most compound pairs sharing little structural overlap and only a smaller subset of close analogues raising the mean, indicating that the potency is not concentrated in a single chemotype but spans multiple structurally distinct series. MCS similarity distribution heatmap provided in Supplementary Figure 9. ZIKV and SARS-CoV followed, with 39 of 426 (9.2%) and 195 of 2,702 (7.2%) compounds below the threshold, respectively, and EBOV was comparable at 21 of 436 (4.8%). The remaining three viruses, which contributed the largest numbers of compounds, showed the lowest potent fractions: LASV 265 of 9,432 (2.8%), MARV 250 of 10,326 (2.4%) and IV 19 of 3,861 (0.5%). Potent fraction was therefore inversely related to dataset size, and the distributions for LASV, MARV and IV were unimodal and centred in the low-micromolar range, consistent with these datasets being dominated by broad phenotypic screening campaigns, whereas the smaller MERS-CoV, ZIKV and EBOV sets are enriched for focused lead-optimization series.

To ask whether potency is associated with particular chemical features, we compared the prevalence of the most common structural fragments and structural alerts between highly potent and potent compounds. Fragment prevalence was broadly similar between the two sets: a benzene ring was present in 87.0% of highly potent versus 91.9% of potent compounds, a trisubstituted nitrogen in 80.9% versus 83.7%, and a carbonyl group in 71.8% versus 67.9% (70.7% versus 66.3% when carboxylic acids were excluded) (Fig. 5b). The one clear exception was the methoxy group, which was substantially enriched among potent compounds (69.5% versus 47.5%). In contrast, four of the five most frequently triggered Brenk alerts (Brenk et al., 2008) were depleted in the highly potent set, most markedly the oxygen–nitrogen single bond (8.9% versus 17.6%) and the imine motif (5.5% versus 13.3%), followed by the nitro group (2.9% versus 6.9%) and the thiocarbonyl group (2.4% versus 6.0%) (Fig. 5c). The aldehyde alert ran counter to this trend, being roughly seven-fold more frequent among highly potent compounds (6.4% versus 0.9%), consistent with a contribution from covalent or covalent-reversible mechanisms within the potent subset. Overall, high potency in this dataset was not associated with gross changes in common scaffold chemistry, but with a modest shift away from reactive or metabolically labile functionality — with the notable exception of aldehydes.

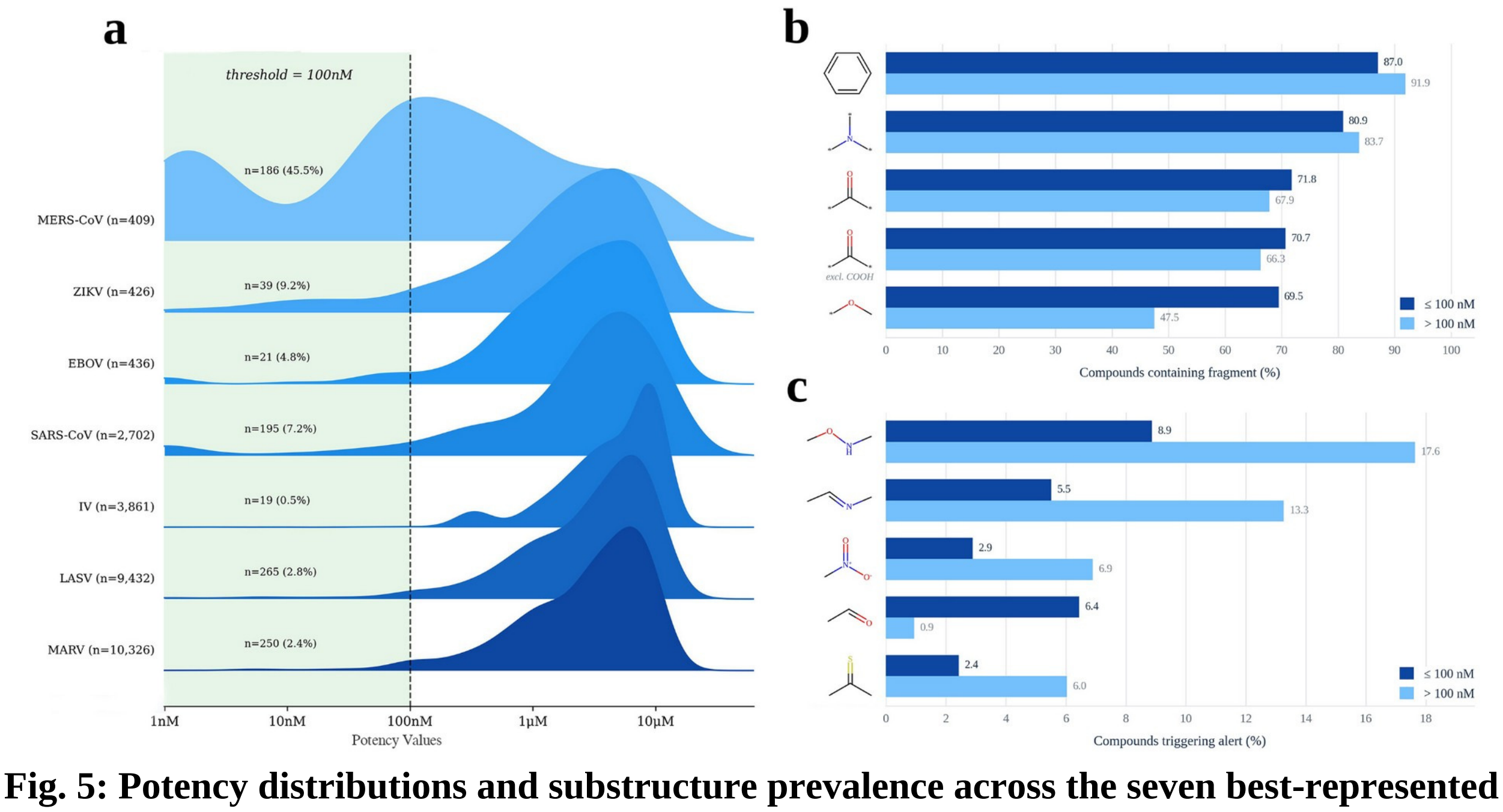

**Fig. 5: Potency distributions and substructure prevalence across the seven best-represented viral targets.**

**a**, Distribution of curated potency values for the seven viruses contributing the largest numbers of molecules, plotted on a $\log_{10}$ concentration axis. The shaded band and dashed vertical line mark the 100 nM potency threshold; values are reported potencies of mixed endpoint (IC50,

EC50, EC90, AC50, Ki, Kd, Potency) and assay type. Each ridge is annotated with the number and percentage of molecules falling below it, ranging from 45.5% (n = 186) for MERS-CoV to 0.5% (n = 19) for IV. Total compound numbers per virus are given in parentheses on the *y*-axis. **b**, Prevalence of the five most common structural fragments among highly potent (≤ 100 nM, upper dark bar) and potent (> 100 nM, lower light bar) compounds, expressed as the percentage of compounds containing each fragment. Fragment structures are shown at left, with wildcard attachment points denoted by asterisks; the fourth entry is the carbonyl fragment with carboxylic acids excluded. **c**, Prevalence of the five most frequently triggered Brenk structural alerts in the same two potency groups. Where a compound had multiple activity measurements against the same virus, the geometric mean was used; otherwise, the reported value was taken directly.

## 3.8. Web interface

The curated library is publicly available, with no registration required, at https://bsadb.denovosciences.ai/ (Fig. 6). Its Home page introduces the Broad-Spectrum Antivirals Research Initiative in four tabs, The Project, Blueprint Viruses, Methodology and Publications, covering the viruses and their druggable targets, the dataset and extraction workflow, and related publications; virus abbreviations link to the corresponding filtered dataset. The Molecular Data page presents the library as a searchable table with virus filters, column descriptions and bulk download (Fig. 6a). Clicking a BSA-DB ID opens a Molecule Details page (Fig. 6b) giving the structure, calculated descriptors, full activity records, targets, assay variability, and links, where available, to ChEMBL, PubChem and BindingDB records. Visual Analysis offers interactive versions of the analyses presented here, and Contact Us accepts data submissions. BSA-DB is maintained by our team and will be updated as the dataset expands. Platform releases follow semantic versioning, with major versions reflecting substantial changes to functionality or data structure and minor versions covering new features and fixes, and all changes are documented in the release notes. The dataset is versioned separately: each update receives a version number shown on the Molecular Data page, and all previous versions remain available for download, so results can be reproduced for any given version. Supporting data retrieved from ChEMBL, PubChem and BindingDB depend on these independently maintained resources and may therefore change over time.

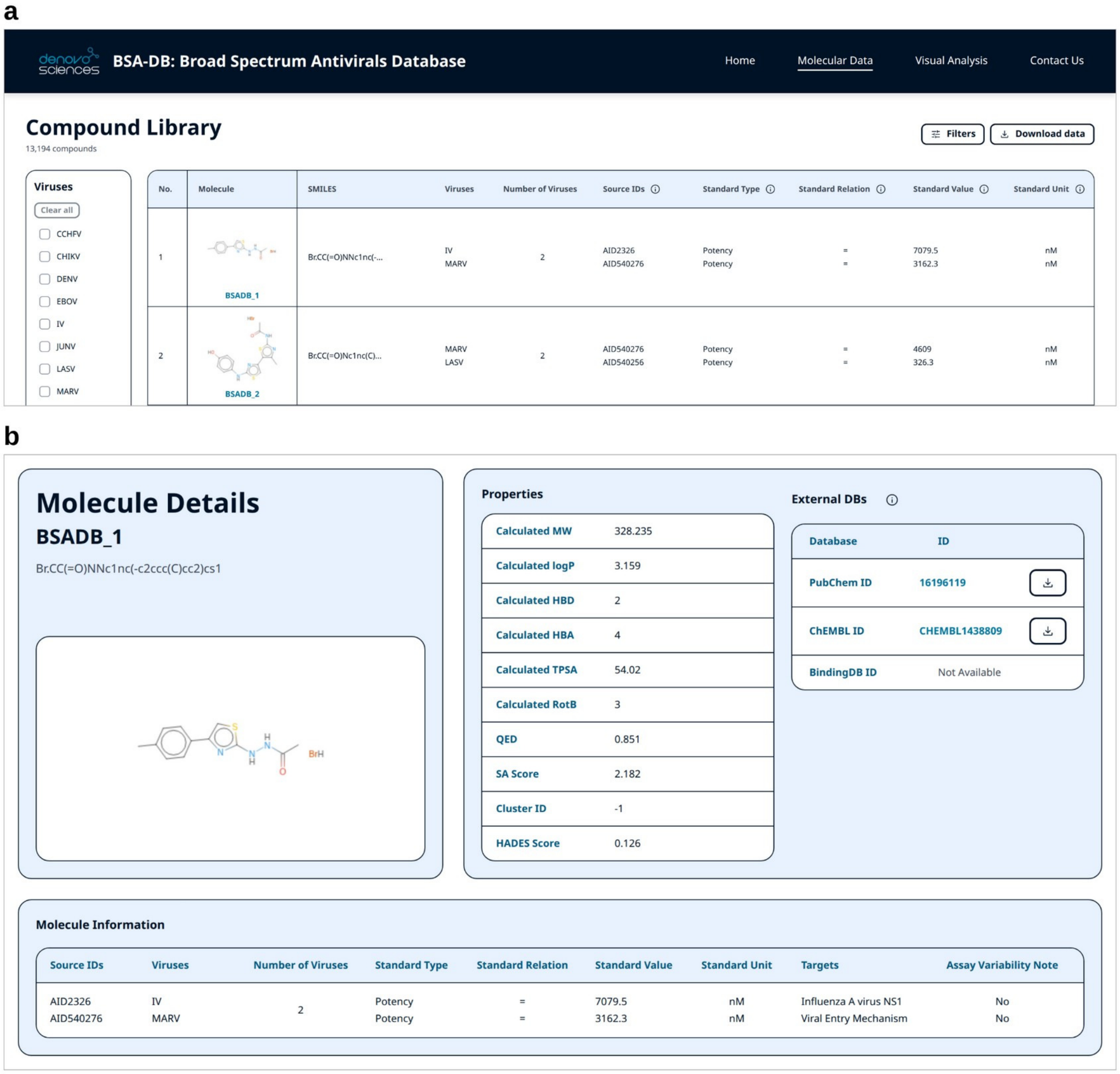

**Fig. 6: The BSA-DB web interface.**

**a**, The Molecular Data page presents the compound library in tabular format, with a filter panel for querying by virus, info icons describing each column, and options to download the dataset in bulk. **b,** Clicking a BSA-DB ID opens a Molecule Details page showing the rendered structure, calculated physicochemical descriptors, cross-references to external databases (ChEMBL, PubChem, BindingDB), and the complete activity record, including targets and assay variability flags.

## 4. Discussion

While public repositories such as PubChem contain millions of bioassay records, their unstructured metadata often obscures cross-viral activity patterns, limiting their direct use for BSA studies without extensive manual curation. Specialized curated databases, in turn, cover only a fraction of the available chemical and bioactivity space because large-scale curation is difficult. For example, the downloadable dataset of DrugVirus.info 2.0 (Ianevski et al., 2022) contains 231 BSAs and 400 BSA-containing combinations, mostly late-stage or approved drugs, whereas AVPdb (Qureshi et al., 2014), DRAVP (Liu et al., 2023), and NA-DB (Luo et al., 2025) focus on specific chemical classes, including peptides and nucleoside analogues. Of these, only DrugVirus.info 2.0 is directly comparable to BSA-DB as a small-molecule resource (Supplementary Method 12). Of its 231 drugs, 96 are present in BSA-DB, leaving 13,063 BSA-DB compounds absent from DrugVirus.info 2.0. The two resources also differ in emphasis: DrugVirus.info 2.0 is broader in viral scope (107 virus species versus 22), whereas BSA-DB samples chemical space far more deeply within the WHO Blueprint priority pathogens. Although their internal chemical diversity is comparable (mean pairwise Tanimoto similarity of 0.121 for BSA-DB and 0.095 for DrugVirus.info 2.0), they occupy chemical space differently. For each DrugVirus.info drug, the most similar BSA-DB compound has a mean Tanimoto similarity of 0.657, indicating that most of these drugs have a close structural counterpart in BSA-DB. Conversely, for each BSA-DB compound, the most similar DrugVirus.info drug reaches a mean of only 0.249, meaning that most BSA-DB compounds have no close counterpart. Thus, although existing databases are valuable, their limited size, chemical coverage, and data scope constrain their use in systematic BSA studies, which calls for a comprehensive resource that integrates antiviral activity across diverse chemical scaffolds, viral targets, and experimental systems.

This study presents BSA-DB, a curated resource comprising 13,194 compounds with reported inhibitory activity (≤10,000 nM) against at least two of 22 viruses prioritized in the WHO R&D Blueprint. BSA-DB was constructed using an LLM-based extraction pipeline that harmonized unstructured metadata from 3,686,843 records across PubChem, ChEMBL, and BindingDB. Benchmarking against expert-curated ChEMBL annotations showed extraction accuracy of 96.15-100% across 20 virus-specific assay sets. Repeated sampling of 1,000 PubChem assays across five independent runs demonstrated high field-level consistency. Manual examination of 200 randomly selected compounds comprising 509 molecule–assay records, a sample size set by Cochran's formula, identified discrepancies in only eight records (1.6%), all attributable to differences in salt forms or stereoisomers rather than incorrect compound identities. These analyses demonstrate that the LLM-based pipeline is highly accurate and reliable for constructing large-scale curated bioactivity databases.

An important feature of BSA-DB is its broad structural diversity. Although the compounds generally occupy physicochemical space compatible with orally bioavailable drug candidates, their underlying structures are highly heterogeneous. The three most abundant Bemis–Murcko scaffolds each account for only 0.24-0.45% of the database, while HDBSCAN identified 497 structural clusters, with a mean Tanimoto similarity below 0.3 within each cluster. Thus, BSA-DB encompasses a broad range of structurally distinct chemical space rather than repeatedly explored antiviral chemotypes.

This diversity provides multiple starting points for identifying and optimizing novel antiviral scaffolds. Researchers can use BSA-DB to identify chemically diverse regions associated with antiviral activity and prioritize both established and less explored chemotypes for further development (Kharatyan et al., 2025). Analysis of privileged antiviral scaffolds further enables systematic investigation of how structural modifications influence antiviral breadth and selectivity. For example, analysis of the S1 scaffold, defined topologically as two six-membered rings connected by a two-atom linker, revealed distinct positional substitution patterns associated with activity against different flaviviruses. Integration of these scaffold-based analyses with larger chemical libraries further expands the searchable chemical space, supporting virtual screening and hit-to-lead optimization of structurally related analogues with potentially improved potency.

BSA-DB may also support the rapid prioritization of existing compounds during emerging viral outbreaks. Structural clustering identified groups with high antiviral breadth, including Vinca alkaloids and camptothecin derivatives, which may represent useful starting points for immediate experimental evaluation against newly emerging or re-emerging viruses.

Despite these advantages, BSA-DB reflects limitations and biases inherent to publicly available bioactivity data. Historical research priorities have resulted in the overrepresentation of intensively studied viruses and comparatively limited information for neglected viruses. Consequently, the apparent density of chemical space associated with some viruses may partly reflect publication and screening bias rather than their true druggability. In addition, extraction and harmonization remain challenging when assay descriptions contain ambiguous viral nomenclature, closely related serotypes, or pseudotyped viral systems. The Lassa-Marburg subset (6,924 compounds) derives almost entirely from two matched PubChem qHTS campaigns that screened one shared library against both viruses (AID540256, AID540276), reflecting screening coverage rather than selective cross-viral activity. Finally, all analyses presented here are retrospective: they describe tendencies across compounds whose activity has already been measured, and their predictive value for new analogues or for viruses beyond the present set has not been experimentally assessed.

## 5. Conclusions

BSA-DB consolidates heterogeneous antiviral bioactivity data into a structured and chemically diverse resource for systematic broad-spectrum antiviral discovery. The database comprises 13,194 compounds with reported activity at or below 10,000 nM against at least two of 22 viruses from the WHO R&D Blueprint priority list, spanning 10 virus families, curated from ChEMBL, PubChem and BindingDB. By combining curated activity data with structural diversity analysis, scaffold-level SAR mapping, and links to expanded analogue space, BSA-DB provides medicinal chemists with actionable starting points for compound prioritization and optimization. Beyond its immediate applications in antiviral discovery and drug repurposing, the multidimensional dataset also provides a foundation for developing machine-learning models to identify unexplored antiviral chemical space and support the design of next-generation broad-spectrum antiviral therapeutics.

## Declaration of competing interests

H.M., N.P., L.K., I.O., A.K., V.H., M.M. and H.Z. are employees of Denovo Sciences Inc., a company engaged in computational drug discovery. BSA-DB is hosted and maintained by Denovo Sciences Inc. To ensure transparency and neutrality, the following measures apply. BSA-DB is compiled exclusively from publicly available databases (ChEMBL, PubChem, and BindingDB), and the curation and analysis procedures are described in full in the Methods and Supplementary Methods, allowing the resource to be independently verified and reconstructed from these public sources. The complete dataset is freely accessible without registration under CC BY-SA 3.0, and may be downloaded in bulk; upstream sources retain their own licenses. No commercial consideration influenced the selection, curation, or presentation of any compound or activity record, and the company asserts no proprietary rights over the aggregated public data.

## CRediT authorship contribution statement

Hovhannes Matevosyan: Conceptualization, Data curation, Methodology, Software, Formal analysis, Visualization, Writing – original draft. Narek Petrosyan: Data curation, Methodology, Validation, Formal analysis, Visualization, Writing – original draft. Levon Kharatyan: Formal analysis, Visualization, Writing – original draft. Iren Okminyan: Software, Visualization, Writing – review and editing. Arman Kharatyan: Investigation, Validation, Writing – review and editing. Astghik Hakobyan: Conceptualization, Investigation, Writing – review and editing. Vardan Harutyunyan: Supervision, Methodology, Writing – review and editing. Mher Matevosyan: Supervision, Methodology, Writing – review and editing. Hovakim Zakaryan: Conceptualization, Supervision, Project administration, Funding acquisition, Writing – original draft, Writing – review and editing. H.M. and N.P. contributed equally to this work.

## Funding

This work was supported by the RA Ministry of Education, Science, Culture and Sports [grant number 23LCG-1F014].

## Data availability

BSA-DB is freely available at https://bsadb.denovosciences.ai/ without registration, with bulk download of the full dataset. Pseudocode and code for scaffold extraction are provided in Supplementary Note 9 and in https://github.com/denovosciencesai/ScaffoldExtractor.

# Supplementary Material

Broad-spectrum antiviral database (BSA-DB): a curated collection of broad-spectrum antiviral agents assembled by large language model mining and expert curation

Hovhannes Matevosyan[1,2], Narek Petrosyan[1,2], Levon Kharatyan[1,2], Iren Okminyan[2], Arman Kharatyan[2], Astghik Hakobyan[1] , Vardan Harutyunyan[2,3], Mher Matevosyan[2,3], Hovakim Zakaryan[1,2,3,*]

1 Laboratory of Antiviral Drug Discovery, Institute of Molecular Biology of NAS, 0014, Yerevan, Armenia
2 Denovo Sciences Inc., 0060, Yerevan, Armenia.
3 Denovo Sciences Doha QFC Branch, Qatar Science and Technology Park, 210000, Doha, Qatar.

*Corresponding author email: hovakimz@denovosciences.ai; h_zakaryan@mb.sci.am

# Supplementary Method 1. Dataset preparation

Highly hazardous viruses associated with a high risk of Public Health Emergencies of International Concern (PHEICs) were selected based on the WHO R&D Blueprint for Epidemics. The selected viruses included Crimean–Congo hemorrhagic fever virus (CCHFV), Chikungunya virus (CHIKV), dengue virus (DENV), Zaire ebolavirus (EBOV), Hantaan virus (HTNV), influenza virus (IV), Junin virus (JUNV), Lassa virus (LASV), Marburg virus (MARV), Middle East respiratory syndrome coronavirus (MERS-CoV), monkeypox virus (MPXV), Nipah virus (NiV), Rift Valley fever virus (RVFV), severe acute respiratory syndrome coronavirus (SARS-CoV), severe fever with thrombocytopenia syndrome virus (SFTSV), smallpox virus (VARV), Sudan ebolavirus (SUDV), tick-borne encephalitis virus (TBEV), vaccinia virus (VACV), Venezuelan equine encephalitis virus (VEEV), West Nile virus (WNV), yellow fever virus (YFV), and Zika virus (ZIKV). Data corresponding to the selected viruses were collected from three publicly available databases: PubChem, ChEMBL, and BindingDB.

Data extraction from PubChem was performed using a multi-step workflow. Initially, virus names (e.g., influenza virus, dengue virus) were used as search queries, and all associated bioassays were retrieved. For each virus, three JavaScript Object Notation (JSON) files—bioassays, substances, and compounds—were downloaded and organized into virus-specific dataframes.

PubChem assay metadata included the assay name, title, and abstract. If the value in the *target name* field explicitly matched a virus name, the assay was directly classified as virus-targeting and stored in a dedicated dataframe. For assays in which the viral target was not explicitly specified, the assay identifier (AID), together with the assay name, title, and abstract, was processed using the DeepSeek Application Programming Interface (API)[S1] — selected for its API availability and cost-effectiveness — to extract structured target information.

Data extraction from ChEMBL and BindingDB followed a similar strategy. In ChEMBL, assays were filtered by selecting *Viruses* under the *Organism Taxonomy L1* category. For each selected virus, assays containing predefined keywords—including virus name synonyms and abbreviations—were retrieved (Supplementary Note 5). Using the ChEMBL API, all available data for molecules tested in these assays, along with their corresponding activity values, were extracted. For BindingDB, the complete dataset was downloaded, and relevant virus-related records were identified and extracted using the same predefined keyword-based approach.

# Supplementary Method 2. LLM text extraction and validation

PubChem text extraction was performed using the DeepSeek API (DeepSeek-V3 version) with a task-specific, tuned prompt. DeepSeek was prompted to extract the *Target Name*, *Organism*, and *Assay Goal Target* following a predefined priority order (assay name, title, and abstract) and a strict set of biological interpretation rules. The complete prompt used for DeepSeek-based extraction is provided in Supplementary Note 6. The extracted fields were appended to the original assay dataframe, and the DeepSeek-annotated assays were subsequently merged with the dataframe containing assays with explicitly defined viral targets.

To evaluate the accuracy of the LLM extraction, a conservative benchmarking strategy was employed. Because the ChEMBL database is manually curated by domain experts, its annotations were used as the ground-truth reference. Virus-related assay records were retrieved from ChEMBL, and the *Description* field of each assay was provided to the DeepSeek model using the prompt described in Supplementary Note 7. The virus identified by the model was then compared with the corresponding *Organism* annotation in ChEMBL to assess the accuracy of the extracted viral information.

Extraction performance was quantified using recall. Recall is defined as:

$$Recall = \frac{TP}{TP + FN}$$

where *TP* and *FN* denote true positives and false negatives, respectively.

To assess the robustness and reproducibility of the LLM-based extraction, a random sample of 1,000 assays from PubChem was selected. Each assay description was submitted to the LLM five independent times using the same prompt (Supplementary Note 6). For each repetition, the model extracted three fields: Target Name, Organism, and Assay Goal Target.

Prior to comparison, extracted values were normalized (lowercased, parenthetical content removed and whitespace collapsed) to avoid penalizing trivial formatting differences. For each field, the five outputs were grouped according to their pairwise similarity (fuzzy string-matching ratio ≥90%; fuzzywuzzy library, https://github.com/seatgeek/fuzzywuzzy), and the largest group was used to define the most common output. The field-level consistency score was calculated as the proportion of valid outputs belonging to this largest group. An overall consistency score was calculated as the mean of the three field-level scores for each assay. A perfect-consistency indicator was assigned when all five runs produced identical values for all three fields.

# Supplementary Method 3. Final data construction and evaluation

After data acquisition from all three databases, several preprocessing and standardization steps were performed. *Target Name* identification was first carried out using assay descriptions from ChEMBL and PubChem, from which standardized target names were extracted using DeepSeek. Since BindingDB provides explicitly defined *Target Names*, only *Target Name* standardization was applied to this dataset. Following target name identification and standardization, the datasets from ChEMBL, PubChem, and BindingDB were merged into a single unified dataset.

Subsequently, all molecular SMILES representations were canonicalized using the RDKit library (version 2025.09.1). The merged dataset was subsequently deduplicated by removing records that were identical across the *SMILES*, *Target Name*, *Standard Type*, *Standard Relation*, *Standard Value*, and *Standard Unit* fields. Activity values were then converted and normalized to nanomolar (nM) units. Values that could not be converted to nM were removed.

Based on the standardized activity values, compounds were grouped by identical SMILES and target name. Within each group, activity consistency was assessed using the reported nM values. Groups containing any activity value greater than 10,000 nM were excluded from further analysis. For the remaining groups (all values ≤ 10,000 nM), variability was evaluated by calculating the ratio of maximum to minimum activity values for exactly the same target. If this ratio was ≥10, the compound was assigned a value of *Yes* in the *Assay Variability Note* column, indicating high intra-group variability; otherwise, it was assigned a value of *No*. When activity data corresponded to different strains or serotypes of the same virus, no additional aggregation or filtering was applied.

Finally, to evaluate the accuracy of the complete data extraction workflow, a random sample of 100 compounds was selected from the final extracted dataset. The extracted information for each compound was manually reviewed against the corresponding source records to verify extraction correctness.

# Supplementary Method 4. Molecular standardization

For further analysis (nucleoside analogues identification, physicochemical and structural characterization, structural clustering, drug-likeness analysis, compounds antiviral potency distributions) all input compounds were standardized prior to analysis using the RDKit (version 2025.09.1) cheminformatics toolkit (https://www.rdkit.org). For each compound, the SMILES representation was parsed into a molecular graph, and the largest covalent fragment was retained to remove counter-ions, salts, and co-crystallized solvents. Formal charges were subsequently neutralized where chemically permissible, and each structure was sanitized to ensure valence consistency and correct aromaticity perception. Stereochemistry was not explicitly processed: it was retained as provided in the source structures (neither enumerated nor removed), and undefined stereocenters were left undefined; RDKit canonicalization preserves specified stereochemistry by default, so stereoisomers reported as distinct structures in the source databases remain distinct entries. Before analysis, compounds with a molecular weight greater than 1000 Da were excluded from the dataset.

# Supplementary Method 5. Nucleoside analogue identification

Nucleoside analogues were identified and validated following the protocol reported by Luo et al.[S2] Initially, candidate nucleoside analogues were identified using 5 natural bases as queries for a substructure search, alongside 9 natural nucleosides and 39 approved nucleoside analogue drugs for a similarity search. Similarity scores were calculated via Tanimoto similarity with Morgan fingerprints (radius = 2, fpsize = 2048), with the similarity threshold set to 0.4 as reported in the original paper. For validation, the identified nucleoside analogues were divided into four groups: "All-match", "Base-match", "Ribose-match", and "invalid nucleic acid analogue", as described in the original paper. While the invalid nucleoside analogues were excluded, those categorized as "All-match", "Base-match", and "Ribose-match" were selected as the final nucleoside analogues. All utilized SMILES strings are listed in Supplementary Note 8. Finally, all analyses were performed using Python and the RDKit library.

## Supplementary Method 6. Physicochemical property calculation and drug-likeness assessment

Eight physicochemical properties were calculated for each compound: molecular weight (MW), lipophilicity (LogP), number of hydrogen bond acceptors (HBA), number of hydrogen bond donors (HBD), number of rotatable bonds, topological polar surface area (TPSA), fraction of $sp^3$ carbons (Fsp3), and number of rings. All properties were computed using standard RDKit descriptors. Distributions of these properties were visualized as histograms with KDE. Druglikeness was assessed using QED and HADES scores. QED integrates eight physicochemical descriptors into a single score, whereas HADES is a soft-voting ensemble of five machine learning models integrating 298 ADMET and physicochemical properties. Score distributions were visualised using KDE plots. Compounds present in the HADES training set were identified by standardised structure and excluded from all score distributions and evaluations. Discrimination of approved compounds from non-approved compounds was quantified by ROC-AUC score. Potency for the priority subset was defined as at least one recorded activity value below 1,000 nM. Structural novelty was quantified as the maximum Tanimoto similarity of each compound to any approved antiviral drug, using Morgan fingerprints (radius 2, 2,048 bits). Murcko scaffolds were computed for the priority subset and grouped into scaffold classes.

## Supplementary Method 7. Ring system and scaffold analysis

Ring systems were extracted from all compounds using RDKit, and benzene was excluded from the analysis to focus on more informative heterocyclic and non-aromatic ring systems. The ten most frequently occurring ring systems were identified and ranked by their percentage occurrence across the dataset.

Murcko scaffolds were extracted from all compounds using the RDKit MurckoDecompose function. The top 100 most abundant scaffolds were selected for further structural analysis. To assess structural diversity and redundancy among these scaffolds, Maximum Common Substructure (MCS)-based pairwise similarity was calculated for all scaffold pairs using the RDKit FindMCS function. MCS similarity is calculated by the equation:

$$MCSsimilarity = \frac{2 \times N_{MCS}}{N_A + N_B}$$

where $N_A$ and $N_B$ are the number of atoms in molecules A and B, and $N_{MCS}$ is the number of atoms of the MCS of molecules A and B. Similarity scores were computed as the ratio of the size of the MCS to the size of the larger of the two compared scaffolds, yielding values between 0 (no common substructure) and 1 (identical scaffolds). The resulting pairwise similarity matrix was visualized as a heatmap.

## Supplementary Method 8. Extraction of generalized scaffolds

To identify promising scaffolds and systematically analyze their chemical space, a multi-step reduction of molecular structures was performed to extract generalized scaffolds. This process started with the reduction of each structure to its Bemis–Murcko scaffold, and the scaffolds were deduplicated. Each unique scaffold was further converted to its cyclic skeleton (CSK) by setting all atoms to carbon and all bonds to single bonds. Pairwise CSK similarity was computed from the MCS similarity. The resulting distances (D = 1 - Sim) were clustered with DBSCAN (ε = 0.1, min_samples = 1, precomputed metric)[S4], grouping related skeletons while allowing singletons. Within each cluster, an MCS was computed for every pair of member skeletons. Pairs with a similarity below 0.9 were discarded, and candidate fragments were screened for broken-ring artifact cases where the MCS severs a ring yet leaves bonds that still require ring membership, which were subsequently rejected. Accepted fragments were pruned to scaffold form by iteratively removing terminal, non-ring atoms of degree ≤ 1, leaving only ring systems and linkers. Identical scaffolds from different pairs were collapsed, keeping the highest-scoring instance. Each cluster's scaffold set was refined by iterative MCS merging. In each iteration, scaffold pairs were merged, and a merged scaffold replaced its two parents only if its similarity was at least as high as both parents and it contained no broken rings; each scaffold merged at most once per iteration, and unmerged scaffolds were carried forward. Merges were pruned and deduplicated, and iteration continued until the set stabilized or a maximum count was reached. Each surviving scaffold was converted into a generalized SMARTS query. The matching source skeletons were identified, and the scaffold was mapped back onto each original molecule, from which per-atom properties (element, aromaticity, ring membership) and per-bond orders were recorded and consolidated: a single observed element was encoded directly, positions matching several elements were generalized to an atomic-number list with aromaticity and ring constraints preserved, and a bond order was kept only when all molecules agreed. The consolidated query was pruned and emitted as SMARTS if it retained more than six atoms; smaller frameworks were discarded as uninformative (Supplementary Figure 10).

All extraction steps were implemented using Python, RDKit, and scikit-learn. The code and pseudocode are accessible in https://github.com/denovosciencesai/ScaffoldExtractor and Supplementary Note 9.

# Supplementary Method 9. Scaffold-based substructure search and chemical space profiling

In order to find promising scaffolds, molecules that had potency values lower than 10 nM for all viruses were separated from the dataset. This resulted in 138 molecules. After that, generalized scaffolds were extracted from these 138 molecules using the algorithm described in the previous section. From these molecules, a total of 19 scaffolds were extracted. Candidate scaffolds were used as substructure queries against the initial viral databases from which the BSA-DB was extracted. For all molecules containing a given scaffold, the MCS similarity was calculated. Molecules with an MCS similarity greater than 0.7 were selected for further analysis. Each retrieved molecule was labelled active or inactive per target according to an activity threshold of 10,000 nM. After searching the database, only 11 scaffolds remained. Two of these were excluded because a more general version was already in the database, leaving a final total of 9 scaffolds. Results are given in Supplementary Table 8.

For every molecule, the representative scaffold was mapped onto the structure by substructure matching, and each scaffold atom position was recorded as an attachment point when it carried a substituent outside the matched core. For a given scaffold, viral target, and scaffold atom position, we tallied the number of active and inactive molecules attached at that position, $a_p$ and $i_p$, and the corresponding class sizes $N_{act}$ and $N_{inact}$ (the number of active and inactive molecules for that scaffold–target pair). To quantify whether an attachment point is associated with activity, we defined a positional activity prevalence metric. This metric normalizes for the sizes of the two classes and is bounded between 0 and 1, where 0 indicates the position is found exclusively in inactive compounds for that target, 1 indicates it is exclusively active, and 0.5 indicates equal representation in both classes.

$$Prevalence = \frac{\frac{a_p}{N_{act}}}{\frac{a_p}{N_{act}} + \frac{i_p}{N_{inact}}}$$

Positions absent from a scaffold–target set were treated as missing (NA). Prevalence values were both visualised as position × target heatmaps (per scaffold) and mapped onto the scaffold structure, colouring each atom by its prevalence.

To analyze the diversity of analogues for the nine selected scaffolds, the PubChem database was explored. The database was accessed on 30 July 2026. Each molecule was matched against the nine scaffolds, and the MCS similarity was calculated for every match. Only molecules exhibiting an MCS similarity greater than 0.7 were retained.

# Supplementary Method 10. Dataset clustering

Compounds were clustered according to structural similarity using molecular fingerprints combined with density-based clustering. Each molecule was encoded as a 2048-bit Morgan fingerprint (radius = 2; ECFP4) using RDKit. Fingerprints were projected into two-dimensional space using Uniform Manifold Approximation and Projection (UMAP) with the Jaccard distance metric[S5]. Clustering was performed using Hierarchical Density-Based Spatial Clustering of Applications with Noise (HDBSCAN). Hyperparameters were optimized by grid search using *min_cluster_size* values of 5, 10, 15, 20, 30, and 50 and *min_samples* values of 1, 3, 5, 10, and 15 (30 parameter combinations). Clustering performance was evaluated using the Density-Based Clustering Validation (DBCV) index, and the parameter combination with the highest DBCV score was selected for downstream analyses. A heatmap of the HDBSCAN hyperparameter grid search results, including the DBCV score, number of clusters, and proportion of noise points, is provided in Supplementary Figure 11.

# Supplementary Method 11. Functional and alerting group analysis

Molecules were categorized by their activity value with a threshold of 100 nM. Then, functional and alerting group prevalence was analyzed in each group. The functional group SMARTS dataset was retrieved from RDKit (https://github.com/rdkit/rdkit/blob/master/Data/FragmentDescriptors.csv), and the alerting group SMARTS were reported by Brenk et al.[S3]

# Supplementary Method 12. Comparison of positional profiles between viral targets

Positional activity prevalence profiles were compared between pairs of viral targets sharing the same scaffold. Profiles were compared between the three Flaviviridae targets represented on scaffold S1 (DENV, WNV and ZIKV). For each pair, the Pearson correlation coefficient was computed across the substituent positions present in both profiles, with positions absent from either treated as missing. Ninety-five per cent confidence intervals were obtained by Fisher z-transformation, in which r is transformed as $z = \operatorname{arctanh}(r)$; the interval is formed as $z \pm 1.96/\sqrt{(n - 3)}$, with n the number of shared positions, and the limits are back-transformed with tanh. P values were corrected for multiple comparisons across the three comparisons by the Benjamini–Hochberg procedure.

# Supplementary Method 13. Comparison with existing antiviral databases

BSA-DB was compared with DrugVirus.info 2.0, AVPdb and DRAVP. AVPdb and DRAVP catalogue antiviral peptides and proteins defined by amino acid sequence and, having no small-molecule structures, are not quantitatively comparable with BSA-DB.

DrugVirus.info 2.0 was compared quantitatively. Its dataset, downloaded on 5 November 2026, comprised 1,518 drug–virus records for 231 unique drugs. Isomeric SMILES were retrieved from PubChem by CID, by name where no CID was given, and for CR-1-31-B from its reported structure (CID 56946209), giving structures for 223 drugs; the remaining 8 are interferons, monoclonal antibodies or large peptides. Structures from both resources were standardised with RDKit (version 2026.03.6) and matched by the first 14 characters of the InChIKey, which encodes connectivity independently of stereochemistry, salt form, and tautomer, or by identical PubChem CID; because BSA-DB stores such variants separately, they were counted once, so that overlap is reported as the number of DrugVirus.info drugs present in BSA-DB. Chemical similarity was quantified with Morgan fingerprints (radius 2, 2,048 bits) and the Tanimoto coefficient. Diversity within each resource was taken as the mean coefficient over all internal compound pairs. Nearest-neighbour similarity was taken as the highest coefficient a compound attains against any compound in the other resource, averaged over all compounds; because the two sets differ greatly in size, these averages are not equivalent, and the calculation was run in both directions, from each DrugVirus.info drug into BSA-DB and from each BSA-DB compound into DrugVirus.info. Viral coverage was compared at the level of virus species and families as annotated in each resource.

# Supplementary References

# Broad-spectrum antiviral database (BSA-DB): a curated collection of broad-spectrum antiviral agents assembled by large language model mining and expert curation

Hovhannes Matevosyan[1,2], Narek Petrosyan[1,2], Levon Kharatyan[1,2], Iren Okminyan[2], Vardan Harutyunyan[2.3], Mher Matevosyan[2.3], Hovakim Zakaryan[1,2,3,*]

1 Laboratory of Antiviral Drug Discovery, Institute of Molecular Biology of NAS, 0014, Yerevan, Armenia
2 Denovo Sciences Inc., 0060, Yerevan, Armenia.
3 Denovo Sciences Doha QFC Branch, Qatar Science and Technology Park, 210000, Doha, Qatar.

*Corresponding author email: hovakimz@denovosciences.ai; h_zakaryan@mb.sci.am

## **Supplementary Note 1.** Benchmarking of DeepSeek extraction

To address the lack of consistent target annotation across the selected viruses, we used DeepSeek to help identify and validate virus-specific bioassays, as described in the Methods section. Before applying DeepSeek to the PubChem dataset, we first evaluated its performance against the ChEMBL database, where curated target and organism annotations are available for comparison.

DeepSeek achieved extraction accuracies ranging from 96.15% to 100% across 20 virus-specific ChEMBL assay sets. However, several limitations were identified. In 0.75% of cases, the model returned "Not specified", indicating that no organism could be identified. DeepSeek also had difficulty distinguishing specific serotypes of IV, SARS-CoV, and DENV, consistent with limitations present in the original database annotations. Additionally, for some EBOV assays involving viral vectors such as HIV or VSV, DeepSeek extracted the vector organism rather than the primary target. Full numerical results of the DeepSeek extraction evaluation, including per-virus accuracies and assay counts, are provided in Supplementary Table 2.

A robustness analysis of DeepSeek extraction was subsequently performed using 1,000 randomly sampled PubChem assays, each processed in five independent runs with an identical prompt. Field-level consistency was high across all three extracted fields. The Target Name field showed an average consistency score of 0.96, with 867 assays (86.7%) showing perfect agreement across all five runs. The Organism field showed a comparable average consistency score of 0.96, with 836 assays (83.6%) in perfect agreement. The Assay Goal Target field showed the highest average consistency score, 0.97, with 892 assays (89.2%) in perfect agreement. However, only 684 assays (68.4%) achieved perfect consistency for Target Name, Organism, and Assay Goal Target together. This gap between the high per-field scores and the lower joint perfect-consistency count reflects that a single field disagreeing in even one of the five runs is sufficient to exclude an assay from the strictest criterion, despite each field individually showing high robustness.

Taken together, the two evaluations indicate that DeepSeek extraction is both accurate against curated reference annotations and stable across repeated runs, with residual errors confined to unresolved serotypes and to viral vectors being reported in place of the primary target.

## **Supplementary Note 2:** PubChem Screening and Analogue Availability for Scaffolds S1–S9.

To evaluate commercial and synthetic analogue space, PubChem was screened against the nine scaffolds using a maximum common substructure (MCS) similarity threshold of 0.7, yielding 1,555,761 total matches (Supplementary Table 9). Analogue representation was exceptionally uneven across the set, spanning six orders of magnitude with a median of 110 molecules per scaffold. A vast majority (99.6%) of all hits were concentrated within just three scaffolds. Conversely, the remaining six scaffolds were sparsely populated: S5 and S7 returned 253 and 110 analogues, respectively; S8, S3, and S2 each yielded fewer than 50; and S9 matched a single PubChem entry.

## **Supplementary Note 3**: Structural relationships among clusters 5, 14, 54 and 2 members.

Cluster 5 (n = 18) comprises nucleoside analogues occupying a remdesivir-like chemical space, modified at the base and sugar moieties to tune activity and metabolic stability. Antiviral coverage was nine viruses for remdesivir (BSADB_1885); five for BSADB_1888; four for BSADB_1200 and GS-441524 (BSADB_10052); three for remdesivir triphosphate (BSADB_10053), BSADB_1201 and sofosbuvir (BSADB_1202); and two for the remaining 11 compounds.

Several members are directly related to one another. Obeldesivir (BSADB_944) is an orally bioavailable prodrug of GS-441524 (BSADB_10052), which is itself the parent nucleoside of remdesivir. Remdesivir triphosphate (BSADB_10053) is the active pharmacological metabolite of remdesivir, and BSADB_1884 is a stereoisomer of remdesivir. Together, these relationships reflect the structural coherence of the cluster around the nucleoside analogue scaffold. Sofosbuvir (BSADB_1202), an approved treatment for hepatitis C virus infection, is structurally more distant from the remdesivir series but nonetheless showed activity against SARS-CoV, ZIKV and YFV, illustrating the cross-virus potential of chemically distinct members within this cluster. Cluster 5 had the second-highest mean antiviral breadth among all structural clusters (2.94) and covered 14 unique viruses, the widest virus coverage of any cluster alongside cluster 14, including CHIKV, EBOV, IV, JUNV, LASV, MARV, MERS-CoV, RVFV, SARS-CoV, YFV, ZIKV, DENV, NiV, and TBEV.

Cluster 14 (n = 15) is built around a carbocyclic nucleoside scaffold. Four compounds — neplanocin A (BSADB_10507), galidesivir (BSADB_10502), BSADB_10491 and BSADB_10511 — were active against four viruses, two (BSADB_10493, BSADB_10492) against three, and the remaining nine against two. Other members include 3-deazaneplanocin (BSADB_10497) and aristeromycin (BSADB_10508). Several of these have been explored clinically as antivirals, supporting cluster 14 as a structurally defined antiviral chemical space.

Cluster 54 showed the highest mean antiviral breadth among all structural clusters (n = 6; mean breadth = 3.17) and is defined by vinca alkaloids, a class of microtubule-targeting natural products and semisynthetic derivatives originally isolated from *Catharanthus roseus* and used in oncology. Vincristine (BSADB_3630) exhibited the broadest antiviral spectrum within the cluster, with activity against ZIKV, LASV, EBOV, and SARS-CoV. Vinorelbine (BSADB_1921) demonstrated activity against MARV, EBOV, and SARS-CoV. Vinblastine (BSADB_3627) was active against LASV, EBOV, and SARS-CoV. Notably, SARS-CoV activity was shared across all six compounds in the cluster. The structural coherence of cluster 54 around the vinca alkaloid scaffold suggests that this chemical class may warrant further investigation as a source of repurposable antiviral agents, particularly against hemorrhagic fever viruses such as EBOV, MARV, and LASV. However, further experimental investigation is required to determine whether the observed activities arise from a shared antiviral mechanism.

Cluster 2 had the third-highest mean antiviral breadth (n = 9; mean breadth = 2.89) and consisted exclusively of camptothecin derivatives. Camptothecins are alkaloids originally isolated from *Camptotheca acuminata* and are established inhibitors of DNA topoisomerase I, with several derivatives approved for cancer treatment. Within this cluster, irinotecan (BSADB_3657) exhibited the broadest activity, against four viruses, while six compounds were active against three viruses and two against two viruses. Several derivatives, including camptothecin (BSADB_3611), gimatecan (BSADB_3606), rubitecan (BSADB_3609), 10-hydroxycamptothecin (BSADB_3610), irinotecan hydrochloride (BSADB_3658), and BSADB_3656, showed activity against multiple combinations of ZIKV, LASV, MARV, EBOV, and SARS-CoV. The high structural similarity of these compounds and their shared activity across multiple viruses make this cluster notable for further investigation of the relationship between the camptothecin scaffold and broad-spectrum antiviral activity.

## **Supplementary Note 4:** Evaluation of Antiviral Coverage and Assay Support for Privileged Scaffolds (MS1–MS3)

To evaluate the breadth of antiviral coverage encoded by these privileged scaffolds, the viruses associated with each scaffold were mapped onto their viral families and broad genome-replication classes, and the assay support underlying each scaffold was quantified. The three scaffolds were collectively active across seven viral families, with MS3 displaying the widest taxonomic reach: its member compounds were active against eight distinct viruses spanning six families (Filoviridae, Orthomyxoviridae, Flaviviridae, Coronaviridae, Togaviridae and Arenaviridae) and three genome-replication classes. MS1 covered six viruses across four families (Filoviridae, Flaviviridae, Coronaviridae and Paramyxoviridae), whereas MS2 was more focused, acting against two viruses from two families (Filoviridae and Arenaviridae); both spanned two genome-replication classes. Importantly, the scaffolds were not artefacts of isolated measurements: they were extracted from diverse, independent bioassays — MS1 from 42 unique assays (14 targets, two activity-metric types), MS3 from 17 assays (14 targets, five metric types) and MS2 from 5 assays (2 targets, a single metric type) — indicating that their broad-spectrum profiles are supported by multiple orthogonal experimental sources. Together, these results reinforce MS1–MS3 as privileged scaffolds and promising starting points for broad-spectrum antiviral drug discovery.

**Supplementary Note 5.** Virus-specific keywords used for assay retrieval

| Selected virus | Predefined retrieval keywords |
|---|---|
| CCHF | CCHF; Crimean Congo; Crimean-Congo; CCHFV; Crimean-Congo hemorrhagic fever virus; Congo fever; Congo hemorrhagic fever |
| CHIKV | Chikungunya; CHIKV; CHIK; chikv; chik |
| Ebola | Ebola; EBOV; EBO; Ebo; EBOLA; Ebola virus - Zaire (1995) |
| Hantaan | Hantaan-virus; Hantaan; HTNV; HNTV; orthohantavirus; Orthohantavirus sp. Hantaan; HFRS virus |
| Lassa | Lassa; LASV; Mammarenavirus Lassa; Arenavirus Lassa; Arenavirus sp. Lassa; LF virus; Lassa hemorrhagic fever virus |
| Marburg | Marburg; MARV; Filovirus Marburg; Filoviridae Marburg virus; Marburg hemorrhagic fever virus; Marburg HF virus; Marburg virus disease; Marburg fever |
| MERS-CoV | MERS-CoV; Middle East Respiratory Syndrome; MERS; Mers |
| Monkeypox | Monkeypox virus; Monkeypox; MPXV; Monkeypox virus strain; Orthopoxvirus monkeypox; Orthopoxvirus sp. Monkeypox; Monkeypox disease; Monkeypox infection; Mpox; Mpox virus |
| Nipah | Nipah virus; Nipah; NiV; Nipah henipavirus; Henipavirus Nipah; Nipah virus strain; Nipah virus infection; Nipah virus disease |
| SARS-CoV | SARS-CoV; SARS virus; SARS coronavirus; Severe Acute Respiratory Syndrome; SARS |
| SFTS | SFTS virus; SFTSV; Severe Fever with Thrombocytopenia Syndrome virus; Severe fever with thrombocytopenia syndrome; SFTS infection; SFTS disease; SFTS virus strain; Phlebovirus SFTSV |
| Smallpox | Smallpox virus; Smallpox; Variola virus; Variola; Variola major; Variola minor; Variola major virus; Variola minor virus; Variola infection; Variola disease |
| Sudan virus | Sudan virus; SUDV; Sudan virus disease; Sudan hemorrhagic fever virus; Sudan HF virus; Filoviridae Sudan virus |
| Venezuelan equine encephalitis virus (VEEV) | Venezuelan equine encephalitis virus; Venezuelan equine encephalitis; VEEV; VEE virus; Venezuelan equine encephalitis virus strain; Alphavirus VEEV; Alphavirus Venezuelan equine encephalitis; Venezuelan equine encephalitis |
| | virus infection |
| Zika | Zika virus; Zika; ZIKV; Zika virus strain; Zika virus infection; Zika virus disease; Flavivirus Zika; Zika flavivirus |
| Rift Valley fever | Rift Valley fever virus; Rift Valley fever; RVFV; Rift Valley fever virus strain; Rift Valley fever virus infection; Rift Valley fever virus disease; Phlebovirus RVFV; Bunyaviridae Rift Valley fever virus |
| Influenza | Influenza virus; Influenza; Flu virus; Flu; Influenza A virus; Influenza B virus; Influenza C virus; Influenza D virus; H1N1; H3N2; H5N1; H9N2 subtype; H2N2 subtype; H7N9 subtype; Avian influenza virus; Swine influenza virus; Human influenza virus; Orthomyxovirus; Orthomyxoviridae influenza virus |
| Yellow fever | Yellow fever virus |
| Junin virus | Mammarenavirus juninense; Junin virus |
| Tick-borne encephalitis | Tick-borne encephalitis; TBE |
| West Nile | West Nile virus |
| Vaccinia virus | Vaccinia virus; Smallpox vaccine |

## **Supplementary Note 6:** DeepSeek API prompt for bioassay extraction from PubChem.

```
Extract and analyze the assay text to provide structured information in JSON format. Follow these guidelines carefully:
EXTRACTION PRIORITY ORDER:
1. First check the "name" field
2. Then check the "title" field
3. Finally check the "abstract" field

FIELD DEFINITIONS:
```

"AID": Extract the exact "aid" value from the JSON data
"Target Name": Identify the specific molecular target (protein, enzyme, receptor, etc.)
"Organism": Identify the organism that the target belongs to (host organism)
"Assay Goal Target": Identify the specific pathogen/virus being tested against in the assay

KEY RULES:
1. For "Target Name": Look for specific protein names (E protein, polymerase, protease, etc.), enzyme names, or receptor names
2. For "Organism": This is typically the HOST organism (human, mouse, hamster cells, etc.) where the target exists
3. For "Assay Goal Target": This is the PATHOGEN being tested (virus name, bacterial species, etc.)
4. If you see phrases like "Antiviral activity against [virus name]" or "Inhibition of [virus name]", then:
- "Assay Goal Target" = [virus name]
- "Organism" = the host cells/organism mentioned
5. If you see "Inhibition of the [protein name] of [virus name]", then:
- "Target Name" = [protein name]
- "Organism" = [virus name] (the organism the protein belongs to)
- "Assay Goal Target" = [virus name]

EXAMPLES:
Example 1: "Antiviral activity against HIV-1 in human T-cells"
AID: [from aid field]
Target Name: "viral replication machinery" (if no specific target mentioned)
Organism: "Homo sapiens"
Assay Goal Target: "HIV-1"

Example 2: "Inhibition of SARS-CoV-2 3CL protease"
AID: [from aid field]
Target Name: "3CL protease"
Organism: "SARS-CoV-2"
Assay Goal Target: "SARS-CoV-2"

Example 3: "Compound activity against influenza virus neuraminidase in MDCK cells"
AID: [from aid field]
Target Name: "neuraminidase"
Organism: "influenza virus"
Assay Goal Target: "influenza virus"

OUTPUT FORMAT:
Return ONLY valid JSON format with these exact keys:
```
{
"AID": "",
 "Target Name": "",
 "Organism": "",
"Assay Goal Target": ""
}
```

If any field cannot be determined, use "Not specified" as the value.

**Supplementary Note 7:** DeepSeek API prompt for bioassay extraction from ChEMBL.

ChEMBL “Organism” extraction prompt

```
prompt = """
Extract the specific target organism from this assay description.
Return ONLY the target organism name in this exact format:
"Target Organism: [extracted target organism name]"
If no clear specific target is mentioned, return:
"Target Organism: Not specified"

Description: {description_text}
"""
```

## Supplementary Note 8: SMILES for Nucleoside Analogues

# NATURAL_BASES
Nc1ncnc2[nH]cnc12
Nc1nc2[nH]cnc2c(=O)[nH]1
Nc1cc[nH]c(=O)n1
O=c1cc[nH]c(=O)[nH]1
Cc1c[nH]c(=O)[nH]c1=O

# NATURAL_NUCLEOSIDES
Nc1ncnc2c1ncn2[C@@H]1O[C@H](CO)[C@@H](O)[C@H]1O
c1nc2c(=O)[nH]c(nc2n1[C@H]3[C@@H]([C@@H]([C@H](O3)CO)O)O)N
O=C1/N=C(/N)\C=C/N1[C@@H]2O[C@@H]([C@@H](O)[C@H]2O)CO
O=C1NC(=O)N(C=C1)[C@@H]2O[C@H](CO)[C@@H](O)[C@H]2O
n2c1c(ncnc1n(c2)[C@@H]3O[C@@H]([C@@H](O)C3)CO)N
c1nc2c(n1[C@H]3C[C@@H]([C@H](O3)CO)O)[nH]c(nc2=O)N
c1cn(c(=O)nc1N)[C@H]2C[C@@H]([C@H](O2)CO)O
Cc1cn(c(=O)[nH]c1=O)[C@H]2C[C@@H]([C@H](O2)CO)O
CC1=CN(C(=O)NC1=O)[C@H]2[C@@H]([C@@H]([C@H](O2)CO)O)O

# NATURAL_RIBOSES
OC[C@H]1O[C@@H](O)[C@H](O)[C@@H]1O
OC[C@H]1O[C@@H](O)C[C@@H]1O

# APPROVED_NA_DRUGS
n3c1c(ncn1[C@H]2/C=C\[C@@H](CO)C2)c(nc3N)NC4CC4
O=C2/N=C(\Nc1n(cnc12)COCCO)N
O=P(O)(O)COCCn1c2ncnc(c2nc1)N
O=C1/N=C(\C=C/N1C[C@H](OCP(=O)(O)O)CO)N
O=C3/N=C\Nc1c3ncn1[C@@H]2O[C@@H](CC2)CO
FC=1\C(=N/C(=O)N(C=1)[C@H]2O[C@H](SC2)CO)\N
C=C1C(CC(C1CO)O)N2C=NC3=C2N=C(NC3=O)N
O=C(OCC(COC(=O)C)CCn1c2nc(ncc2nc1)N)C
O=C2/N=C(\Nc1n(cnc12)COC(CO)CO)N
O=C(OCC(OCn1c2N\C(=N/C(=O)c2nc1)N)CO)[C@@H](N)C(C)C
O=C1/N=C(/N)\C=C/N1[C@@H]2O[C@@H](SC2)CO
CCC(COC(=O)[C@@H](NP(=O)(Oc1ccccc1)OC[C@H]1O[C@@]([C@@H]([C@@H]1O)O)(C#N)c1ccc2n1ncnc2N)C)CC
C[C@@H](C(=O)OC(C)C)N[P@](=O)(OC[C@@H]1[C@H]([C@@]([C@@H](O1)N2C=CC(=O)NC2=O)(C)F)O)OC3=CC=CC=C3
O=C1/C(=C\N(C(=O)N1)[C@@H]/2O[C@@H](\C=C\2)CO)C
Cc1cn([C@@H]2C[C@@H](O)[C@H](CO)O2)c(=O)[nH]c1=O
O=P(O)(O)CO[C@H](C)Cn1c2ncnc(c2nc1)N
O=C(OCCOCn1c2N\C(=N/C(=O)c2nc1)N)[C@@H](N)C(C)C

O=C1NC(C(C)=CN1[C@@H]2O[C@H](CO)[C@@H](N=[N+]=[N-])C2)=O
OC[C@@H](O1)[C@@H](O)[C@@H](O)[C@@H]1N2N=C(C(N)=O)N=C2
FC(F)(F)C=1C(=O)NC(=O)N(C=1)[C@@H]2O[C@@H]([C@@H](O)C2)CO
O=C1/N=C(/N)\C=C/N1[C@@H]2O[C@@H](CC2)CO
O=C2/N=C(\Nc1n(cnc12)CCC(CO)CO)N
I\C1=C\N(C(=O)NC1=O)C2C[C@H](O)[C@@H](CO)O2
C1=NC(=C2C(=N1)N(C=N2)[C@H]3[C@H]([C@@H]([C@H](O3)CO)O)O)N
Br[C@H]=CC=1C(=O)NC(=O)N(C=1)[C@@H]2O[C@@H]([C@@H](O)C2)CO
O=C/1NC(=O)N(\C=C\1C)[C@H]2O[C@H]([C@H](O)[C@H]2F)CO
n1c(nn(c1)[C@@H]2O[C@@H]([C@@H](O)[C@H]2O)CO)C(=[N@H])N
CC(C)C(=O)OC[C@H]1O[C@@H](n2ccc(NO)nc2=O)[C@H](O)[C@@H]1OO[C@H]3[C@H]
(c2c1ncnc(N)c1[nH]c2)N[C@H](CO)[C@H]3O
O=C1/N=C(/N)\C=C/N1[C@@H]2O[C@@H]([C@@H](O)[C@@H]2O)CO
c1cn(c(=O)nc1N)[C@H]2C([C@@H]([C@H](O2)CO)O)(F)F
FC=1\C(=N/C(=O)N(C=1)[C@@H]2O[C@@H]([C@@H](O)[C@H]2O)C)\NC(=O)OCCCCC
FC=1C(=O)NC(=O)N(C=1)[C@@H]2O[C@@H]([C@@H](O)C2)CO
O=C1/N=C(\N=C/N1[C@@H]2O[C@@H]([C@@H](O)[C@H]2O)CO)N
O=C1/N=C(\N=C/N1[C@@H]2O[C@@H]([C@@H](O)C2)CO)N
Fc1nc(c2ncn(c2n1)[C@@H]3O[C@@H]([C@@H](O)[C@@H]3O)CO)N
Clc1nc(c2ncn(c2n1)[C@@H]3O[C@@H]([C@@H](O)C3)CO)N
Clc1nc(c2ncn(c2n1)[C@@H]3O[C@@H]([C@@H](O)[C@@H]3F)CO)N
n2c1c(nc(nc1OC)N)n(c2)[C@@H]3O[C@@H]([C@@H](O)[C@@H]3O)CO

## Supplementary Note 9: Pseudocode for Extraction of Generalized Scaffolds

**Algorithm** | Generalized scaffold extraction

```
Input: compound set M; ε = 0.1, τ = 0.9, n_min = 6
Output: set Q of generalized SMARTS queries
procedure GenScaffolds(M)
    B ← unique{Murcko(m) : m ∈ M}
    S ← unique{Csk(b) : b ∈ B}
    D_ij ← 1 − sim(S_i, S_j)
    C ← Dbscan(D, ε, minPts = 1)
    Q ← ∅
    for c ∈ C do
        F ← ∅
        for (x, y) ∈ pairs(c) do
            f ← Trim(Mcs(x, y)),   s ← sim(x, y)
            if s ≥ τ and ¬ Severed(f) then
                F ← F ∪ {(f, s)}
        repeat
            for (a, b) ∈ pairs(F) do                ▷ one merge per round
                m ← Trim(Mcs(a, b))
                if sim(a, b) ≥ max(s_a, s_b) and ¬ Severed(m) then
                    F ← (F \ {a, b}) ∪ {m}
        until F unchanged
        for f ∈ F do
            q ← Trim(Generalize(f, c, M))
            if |q| > n_min then
                Q ← Q ∪ {Smarts(q)}
    return Q
```

**Algorithm | Generalized scaffold extraction.** Pseudocode for the extraction of generalized scaffold queries. Block structure is indicated by indentation. Operators: **Murcko**, reduction to the Bemis–Murcko scaffold; **Csk**, reduction to the cyclic skeleton, in which every atom is set to carbon and every bond to single; **Mcs**, maximum common substructure of two structures; sim(*x*, *y*) = |**Mcs**(*x*, *y*)| / max(|*x*|, |*y*|), where |·| denotes the heavy-atom count; **Dbscan**, density-based clustering of the precomputed distance matrix *D*; **Trim**, iterative deletion of non-ring atoms of degree ≤ 1, leaving ring systems and their linkers; **Severed**, rejection of fragments in which the maximum common substructure cuts a ring while leaving atoms or bonds that still require ring membership; **Generalize**, consolidation of element, aromaticity, ring membership and bond order observed at each position across all matching compounds, encoding a single observed element directly and several as an atomic-number list, and retaining a bond order only where all compounds agree; **Smarts**, emission of the resulting query in SMARTS notation. *s*<sub>a</sub> and *s*<sub>b</sub> denote the similarity scores stored with frameworks *a* and *b*. The annotation "one merge per round" marks the constraint that each framework may take part in at most one merge within a single pass of the **repeat** loop: once *a* and *b* have been merged, both are excluded from further pairing until the next round, so that every round consists of disjoint pairwise merges and no framework can be

consumed by two children simultaneously. Frameworks left unpaired in a round are carried forward unchanged, and the loop terminates once the set of frameworks no longer changes.

**Supplementary Table 11. Sensitivity of the priority subset to the size of the screening budget (N).**

| N | HADES cutoff | Elite (n) | Elite (%) | Viruses | Unique scaffolds | MS1 (n) | MS2 (n) | MS3 (n) |
|---|---|---|---|---|---|---|---|---|
| 50 | 0.957 | 45 | 90 | 11 | 41 | 3 | 0 | 4 |
| 75 | 0.934 | 62 | 82.7 | 11 | 60 | 5 | 0 | 4 |
| 100 | 0.906 | 83 | 83 | 13 | 80 | 7 | 1 | 4 |
| 125 | 0.88 | 103 | 82.4 | 14 | 93 | 9 | 3 | 4 |
| 150 | 0.834 | 127 | 84.7 | 15 | 112 | 9 | 4 | 4 |
| 175 | 0.805 | 150 | 85.7 | 17 | 131 | 9 | 4 | 4 |
| 200 | 0.765 | 175 | 87.5 | 18 | 151 | 9 | 4 | 4 |

**Supplementary Figure 1**

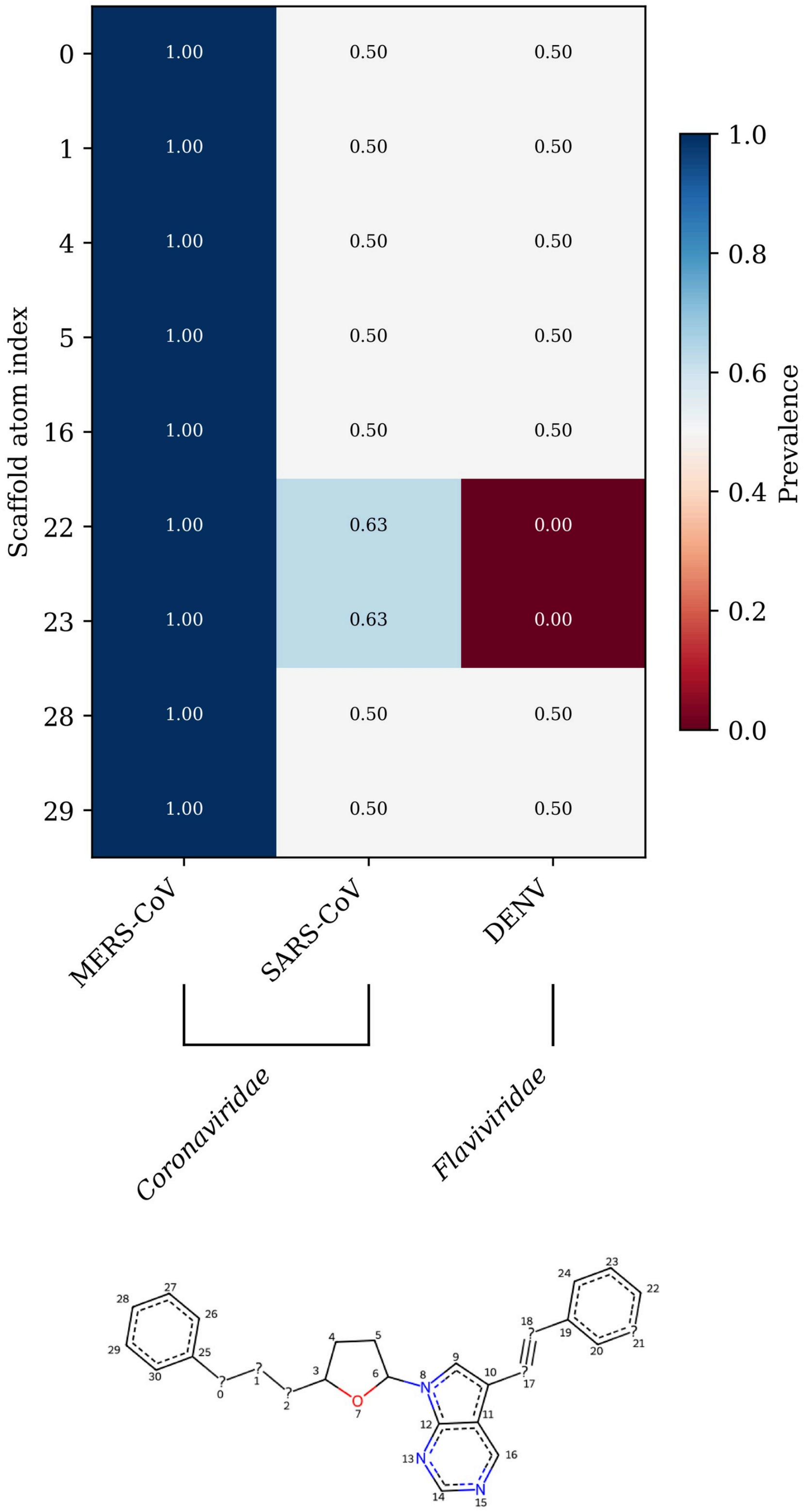

Generalized Scaffold S2

**Supplementary Figure 2**

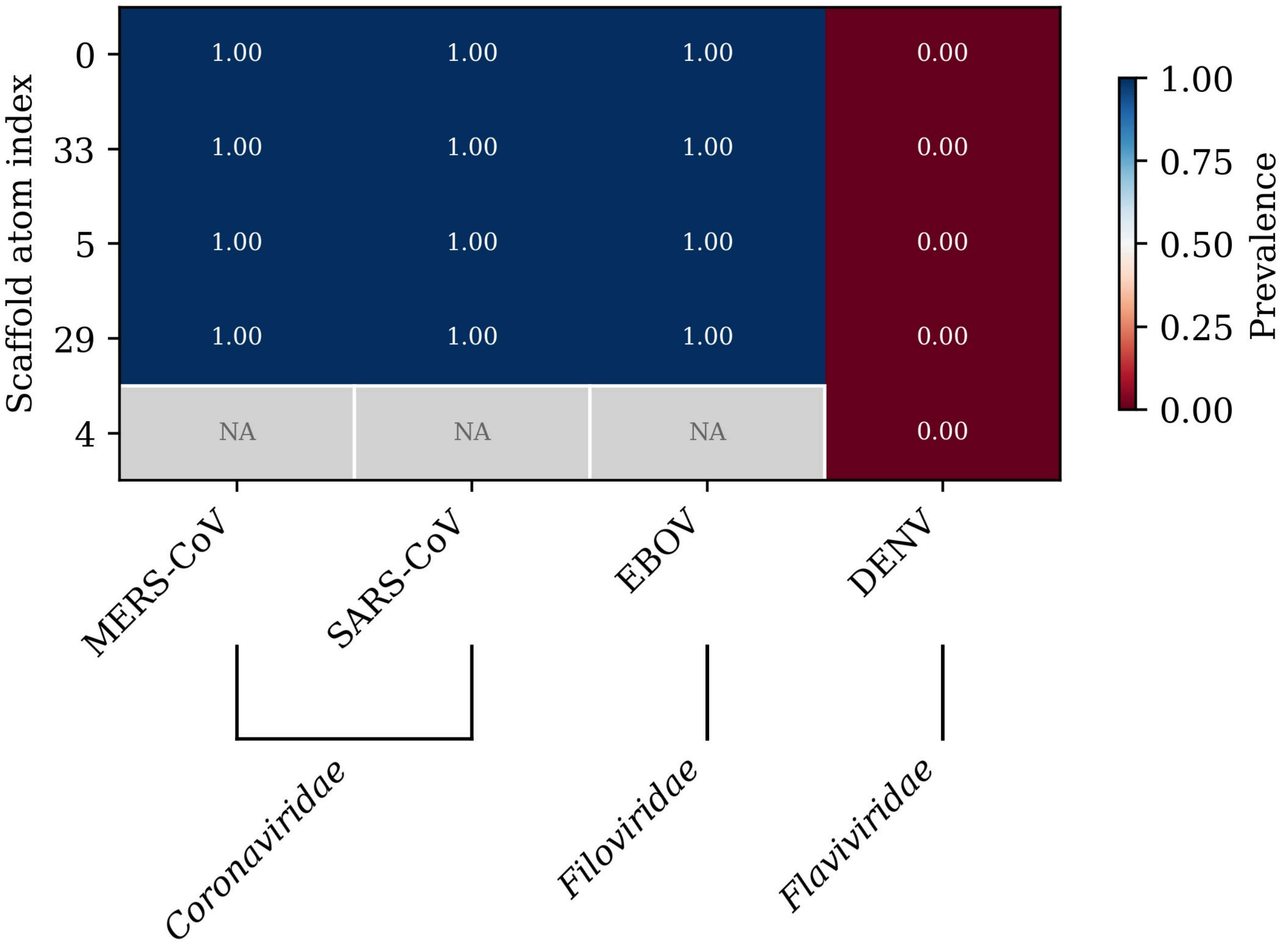

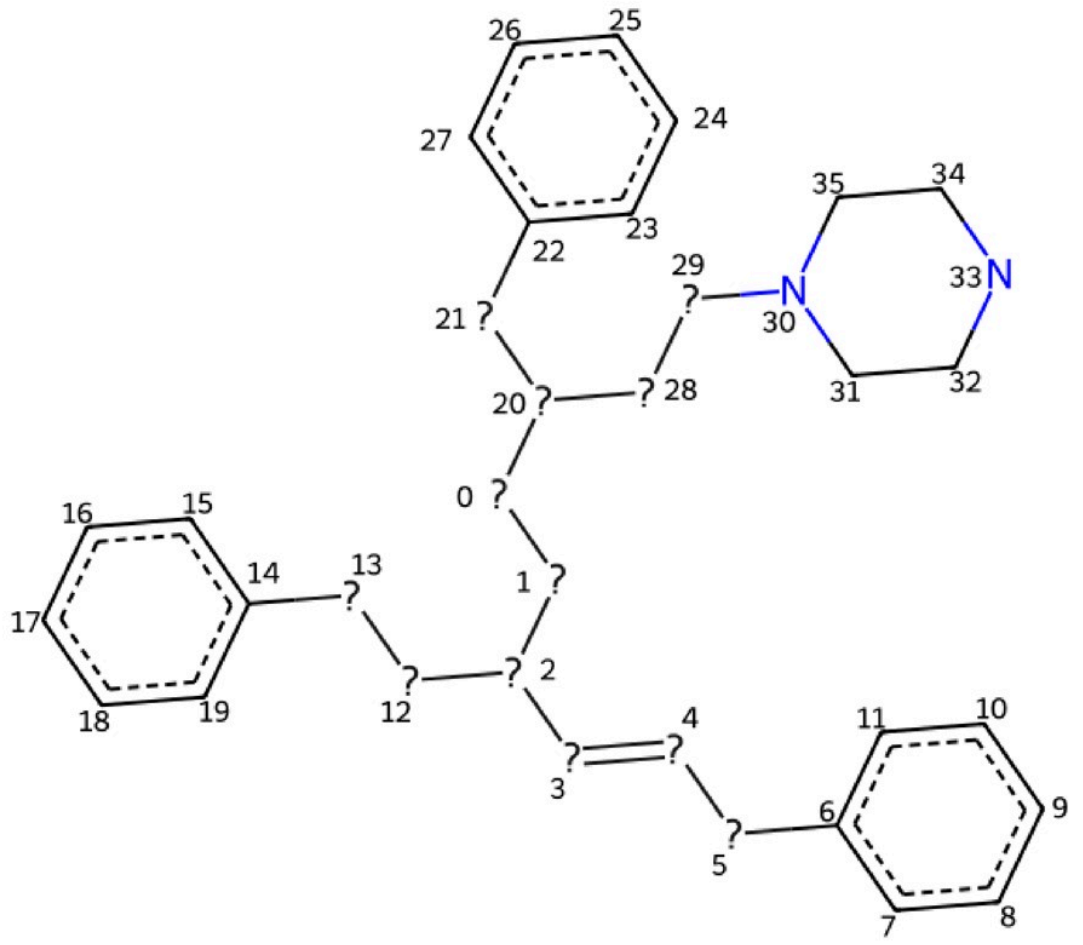

Generalized Scaffold S3

## Supplementary Figure 3

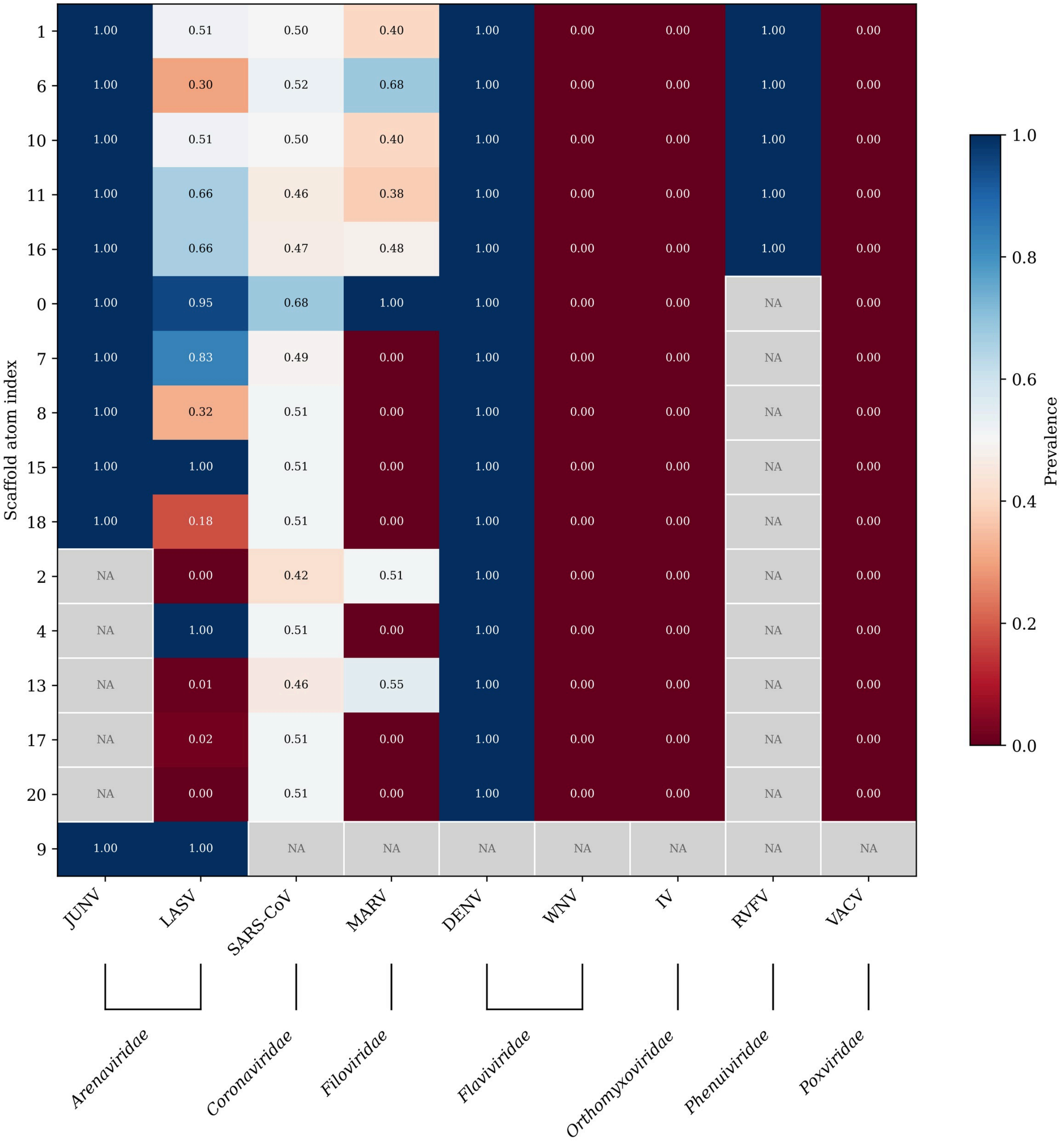

| Scaffold atom index | JUNV | LASV | SARS-CoV | MARV | DENV | WNV | IV | RVFV | VACV |
|---|---|---|---|---|---|---|---|---|---|
| 1 | 1.00 | 0.51 | 0.50 | 0.40 | 1.00 | 0.00 | 0.00 | 1.00 | 0.00 |
| 6 | 1.00 | 0.30 | 0.52 | 0.68 | 1.00 | 0.00 | 0.00 | 1.00 | 0.00 |
| 10 | 1.00 | 0.51 | 0.50 | 0.40 | 1.00 | 0.00 | 0.00 | 1.00 | 0.00 |
| 11 | 1.00 | 0.66 | 0.46 | 0.38 | 1.00 | 0.00 | 0.00 | 1.00 | 0.00 |
| 16 | 1.00 | 0.66 | 0.47 | 0.48 | 1.00 | 0.00 | 0.00 | 1.00 | 0.00 |
| 0 | 1.00 | 0.95 | 0.68 | 1.00 | 1.00 | 0.00 | 0.00 | NA | 0.00 |
| 7 | 1.00 | 0.83 | 0.49 | 0.00 | 1.00 | 0.00 | 0.00 | NA | 0.00 |
| 8 | 1.00 | 0.32 | 0.51 | 0.00 | 1.00 | 0.00 | 0.00 | NA | 0.00 |
| 15 | 1.00 | 1.00 | 0.51 | 0.00 | 1.00 | 0.00 | 0.00 | NA | 0.00 |
| 18 | 1.00 | 0.18 | 0.51 | 0.00 | 1.00 | 0.00 | 0.00 | NA | 0.00 |
| 2 | NA | 0.00 | 0.42 | 0.51 | 1.00 | 0.00 | 0.00 | NA | 0.00 |
| 4 | NA | 1.00 | 0.51 | 0.00 | 1.00 | 0.00 | 0.00 | NA | 0.00 |
| 13 | NA | 0.01 | 0.46 | 0.55 | 1.00 | 0.00 | 0.00 | NA | 0.00 |
| 17 | NA | 0.02 | 0.51 | 0.00 | 1.00 | 0.00 | 0.00 | NA | 0.00 |
| 20 | NA | 0.00 | 0.51 | 0.00 | 1.00 | 0.00 | 0.00 | NA | 0.00 |
| 9 | 1.00 | 1.00 | NA | NA | NA | NA | NA | NA | NA |

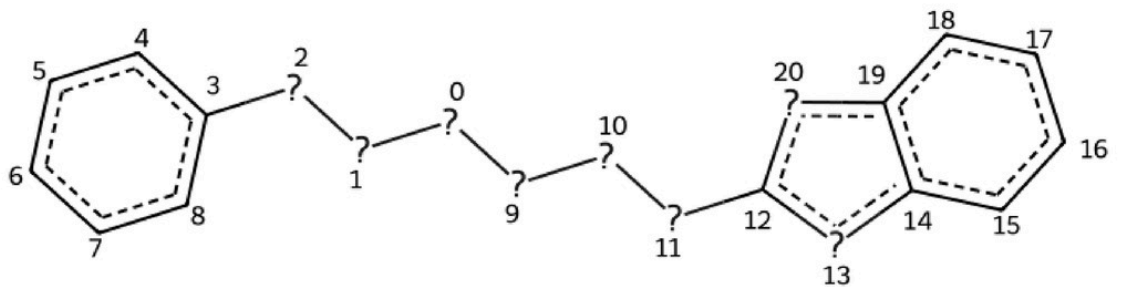

Generalized Scaffold S4

**Supplementary Figure 4**

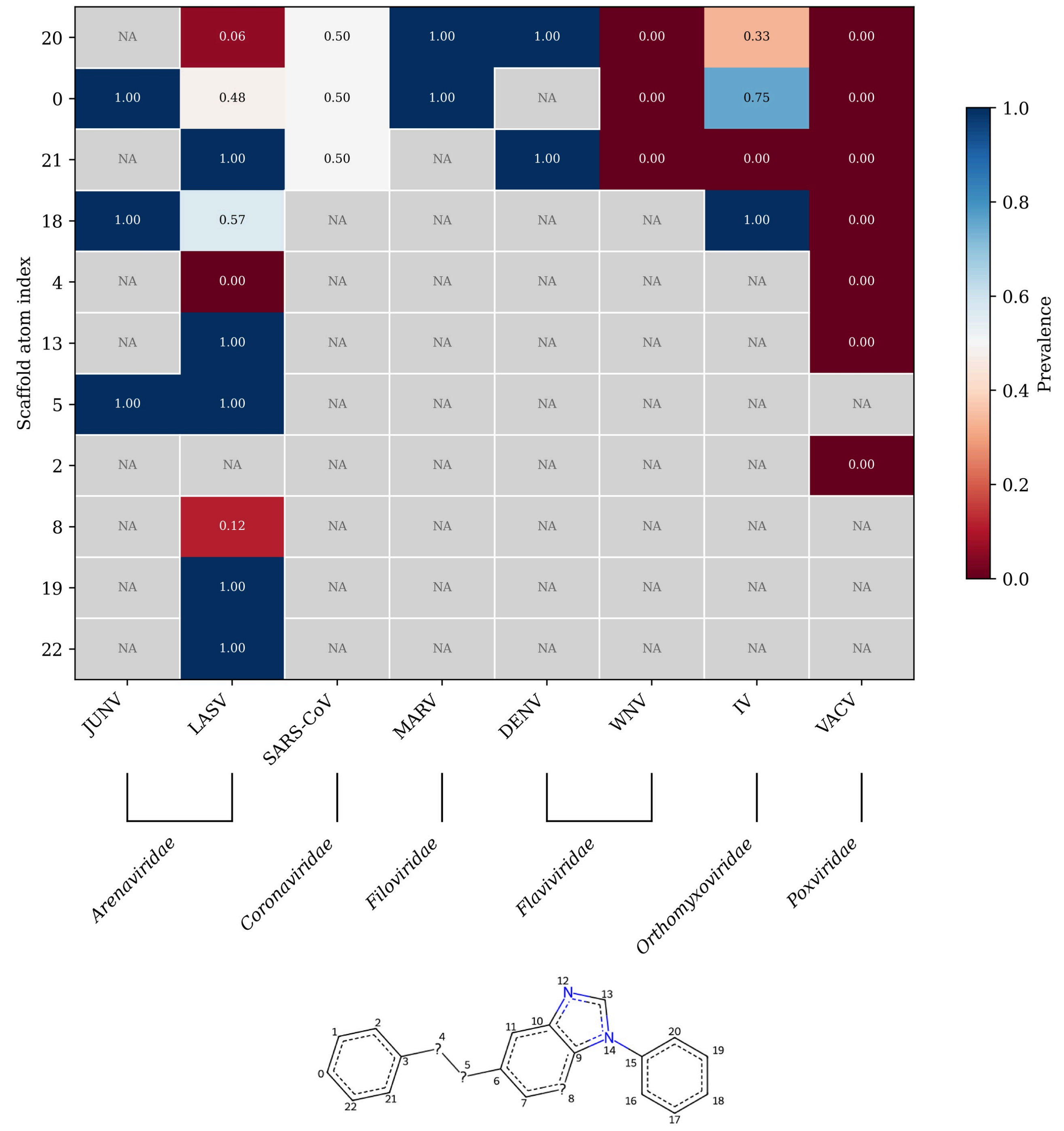

Generalized Scaffold S5

## Supplementary Figure 5

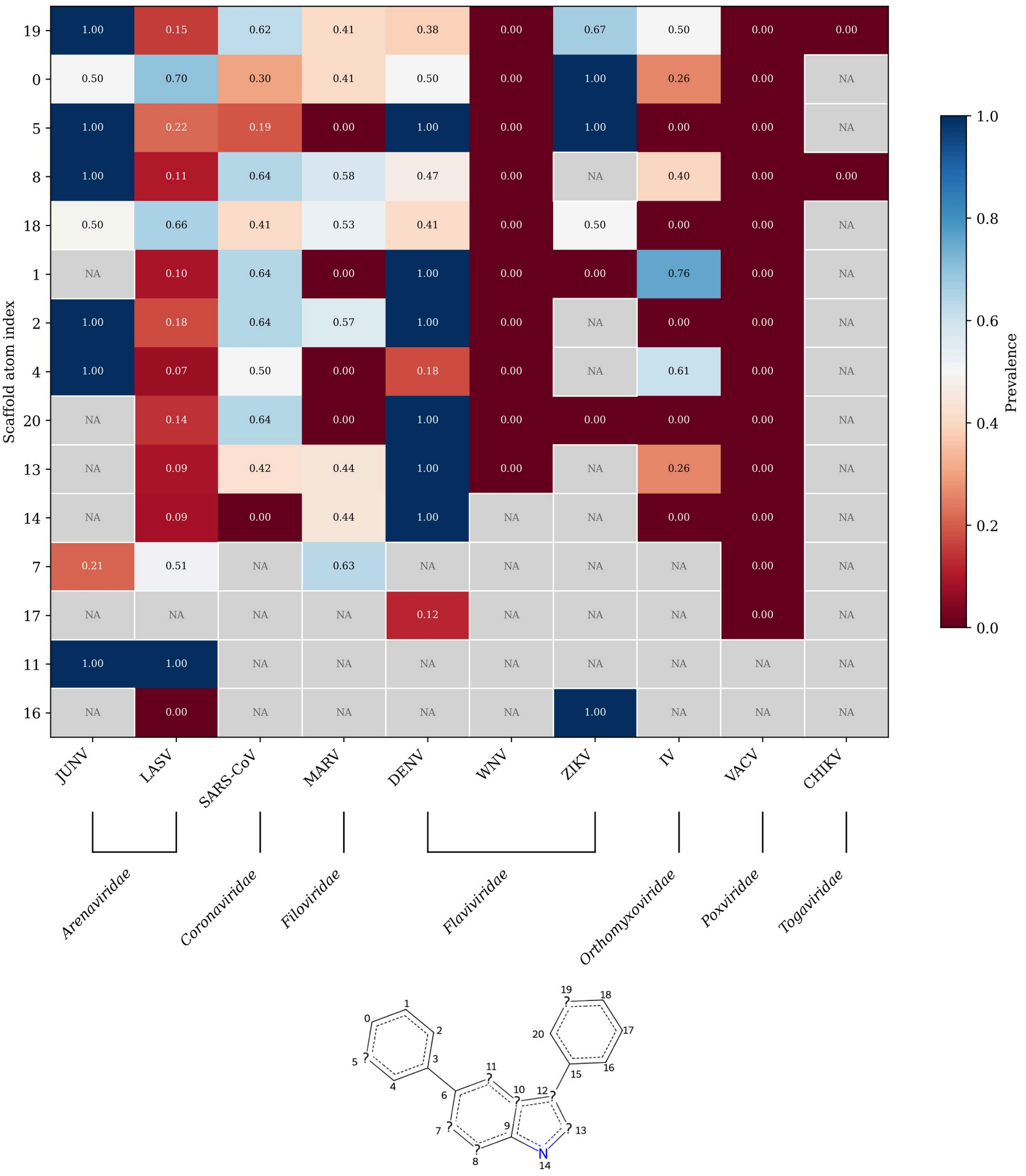

Generalized Scaffold S6

**Supplementary Figure 6**

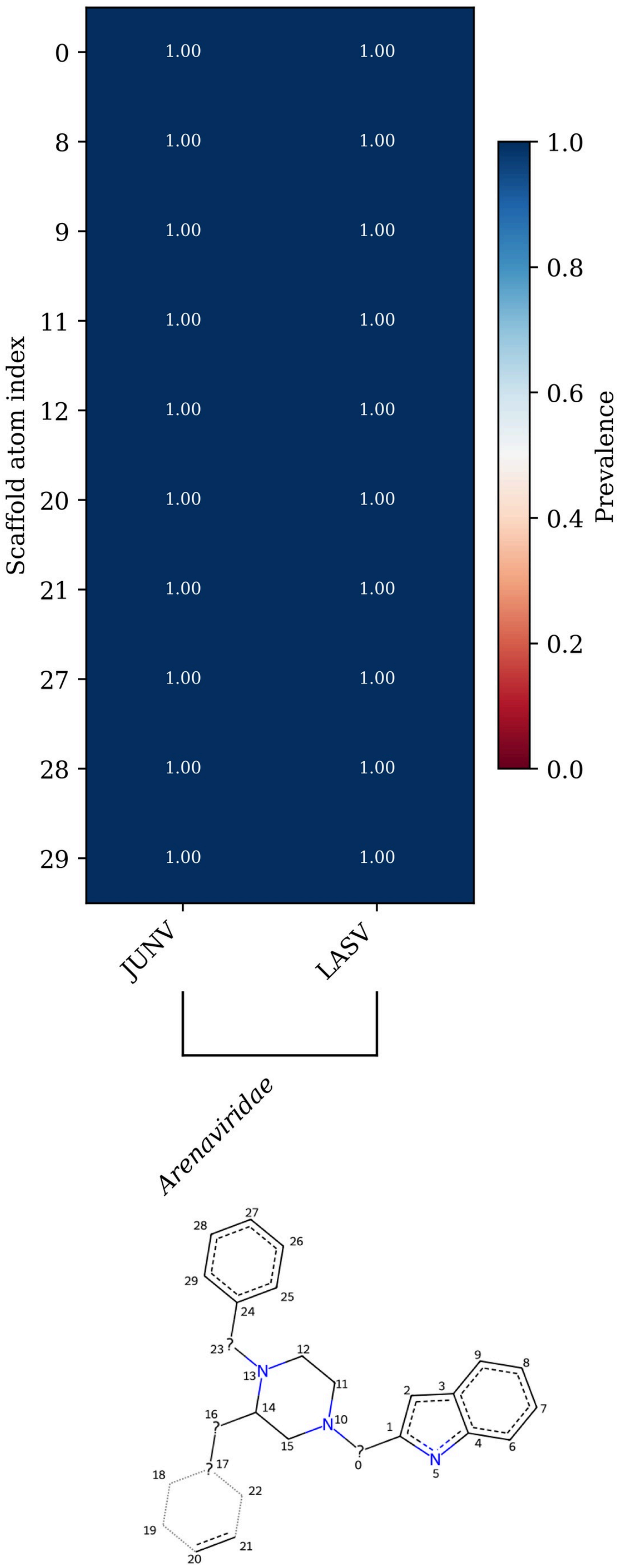

Generalized Scaffold S7

**Supplementary Figure 7**

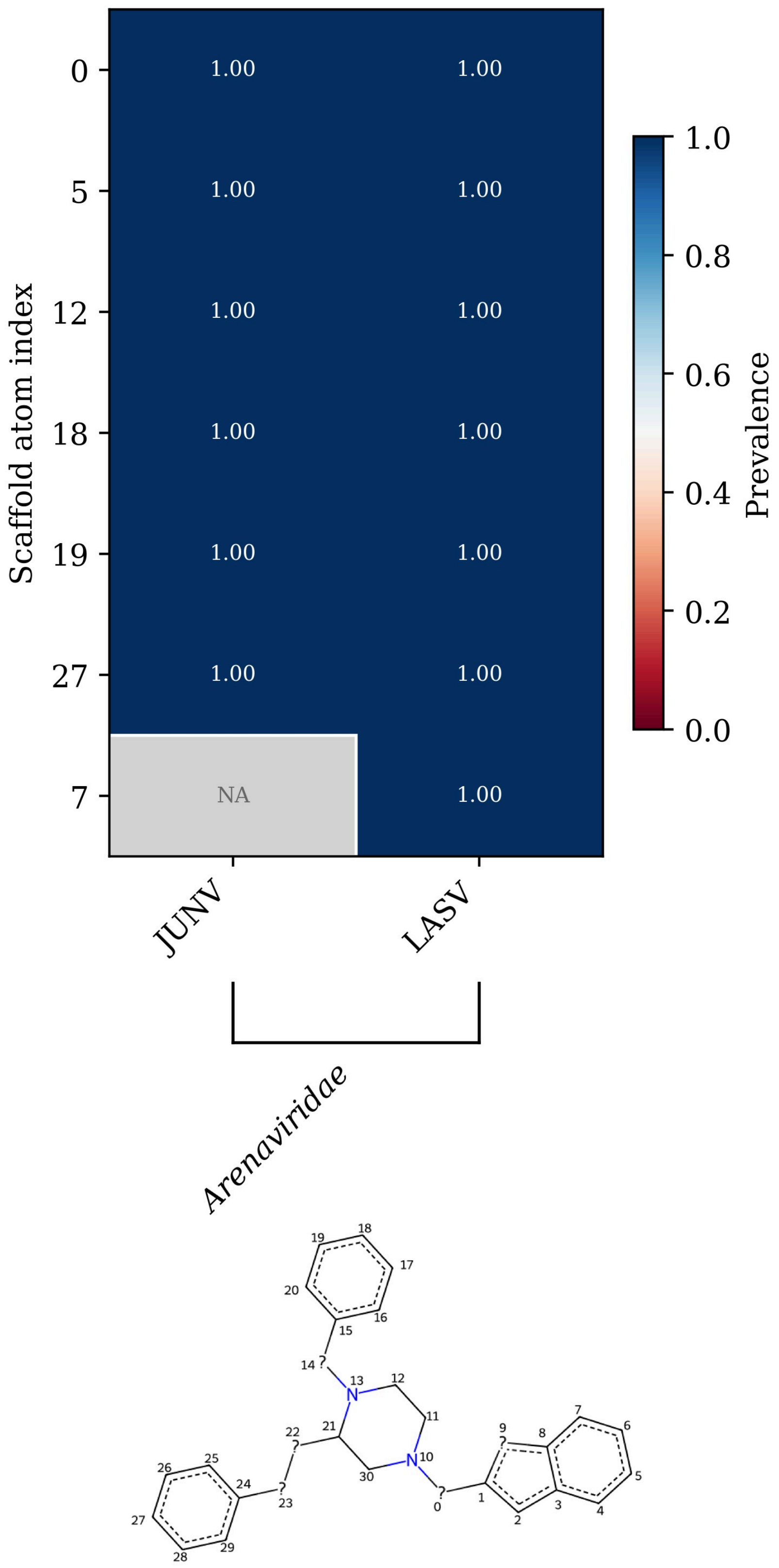

Generalized Scaffold S8

**Supplementary Figure 8**

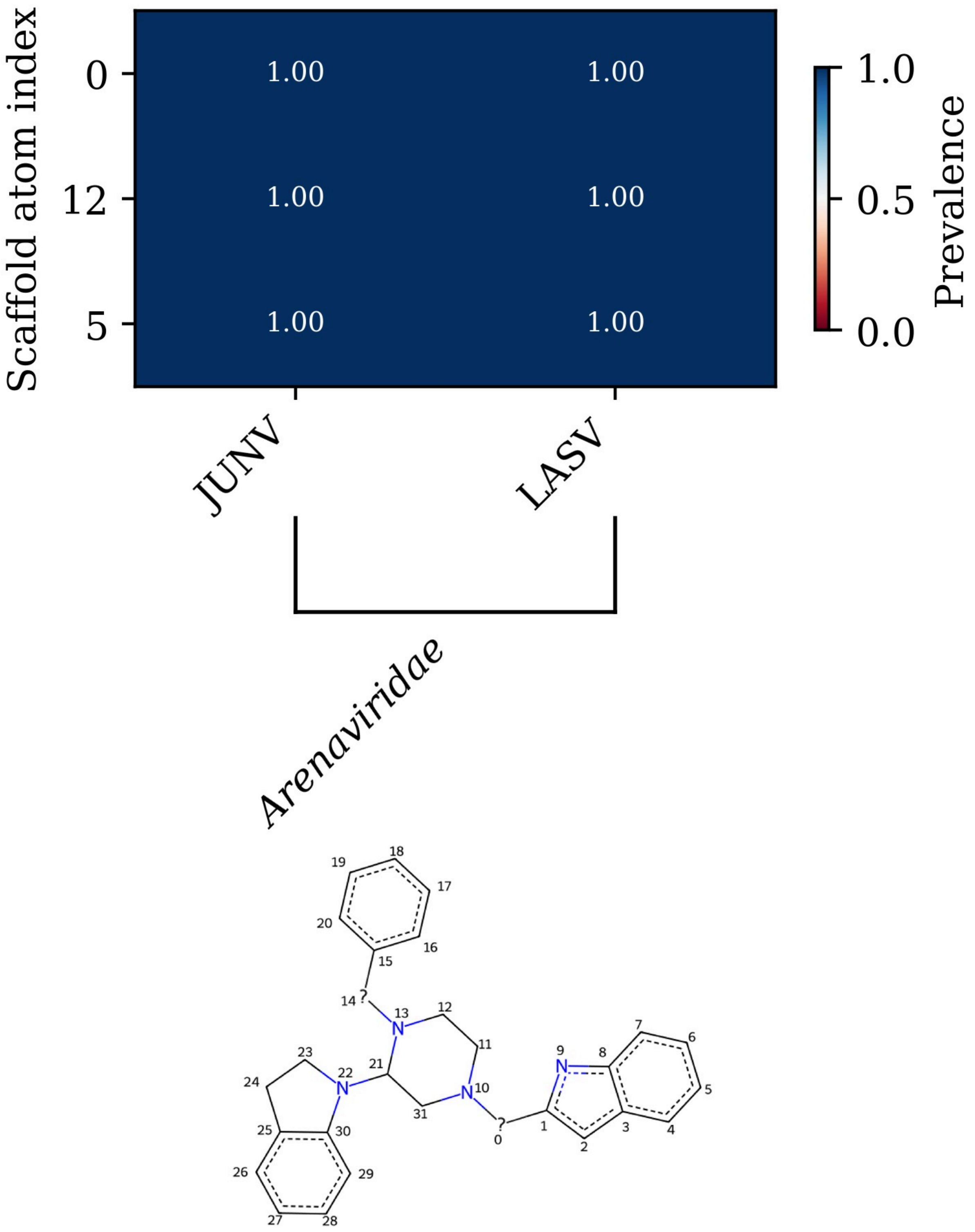

Generalized Scaffold S9

## Supplementary Figure 9

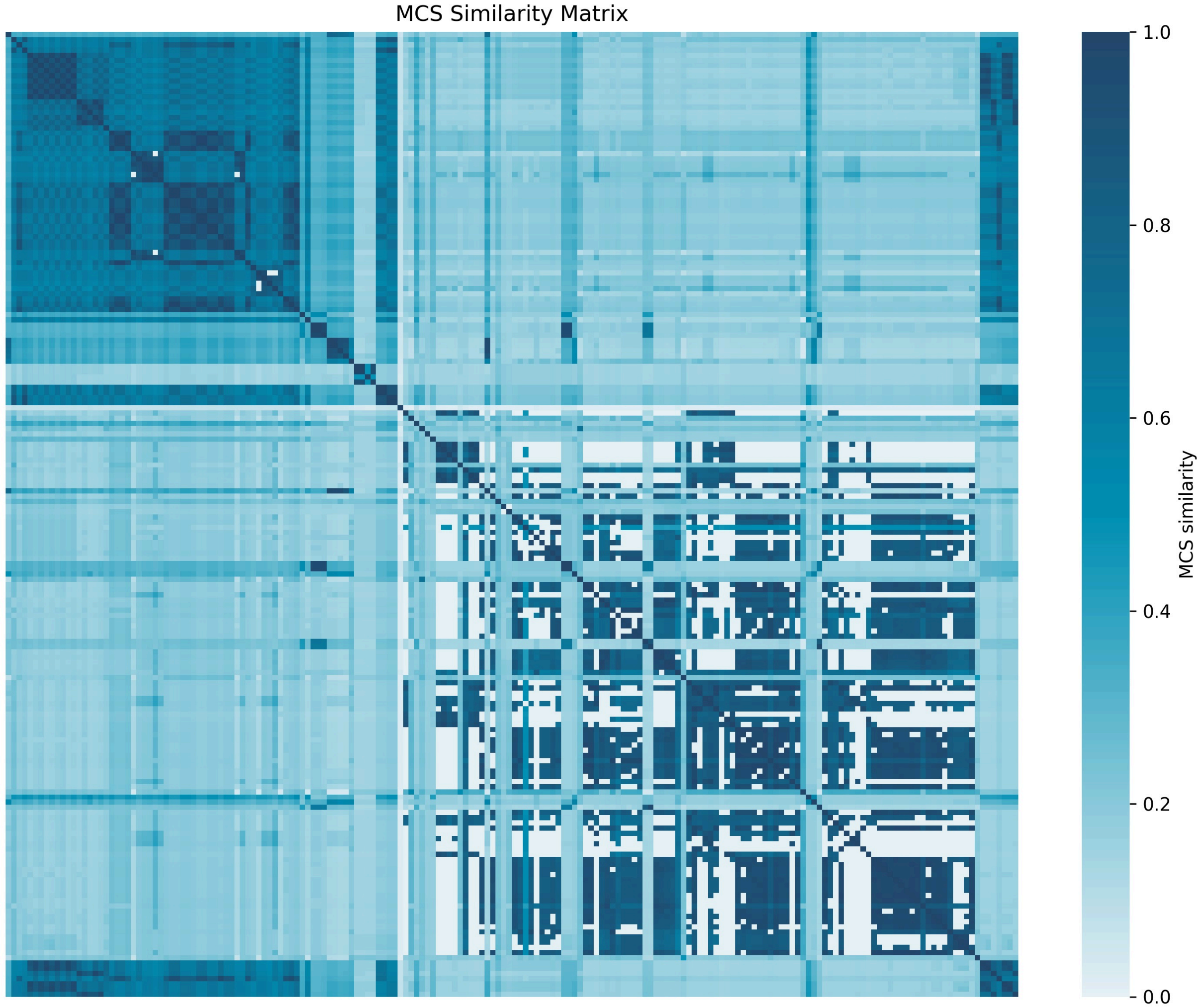

## Supplementary Figure 10

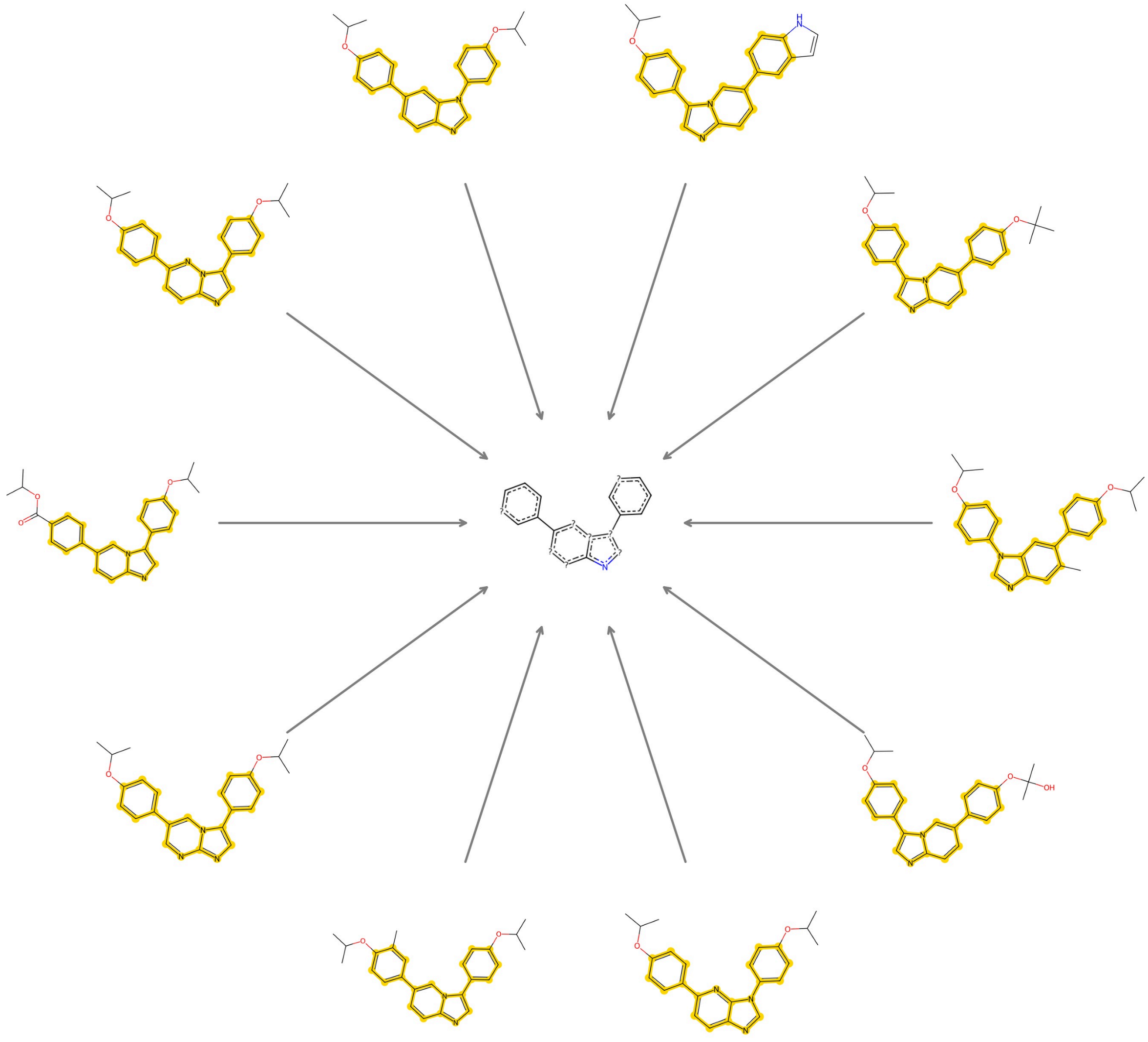

# Supplementary Figure 11

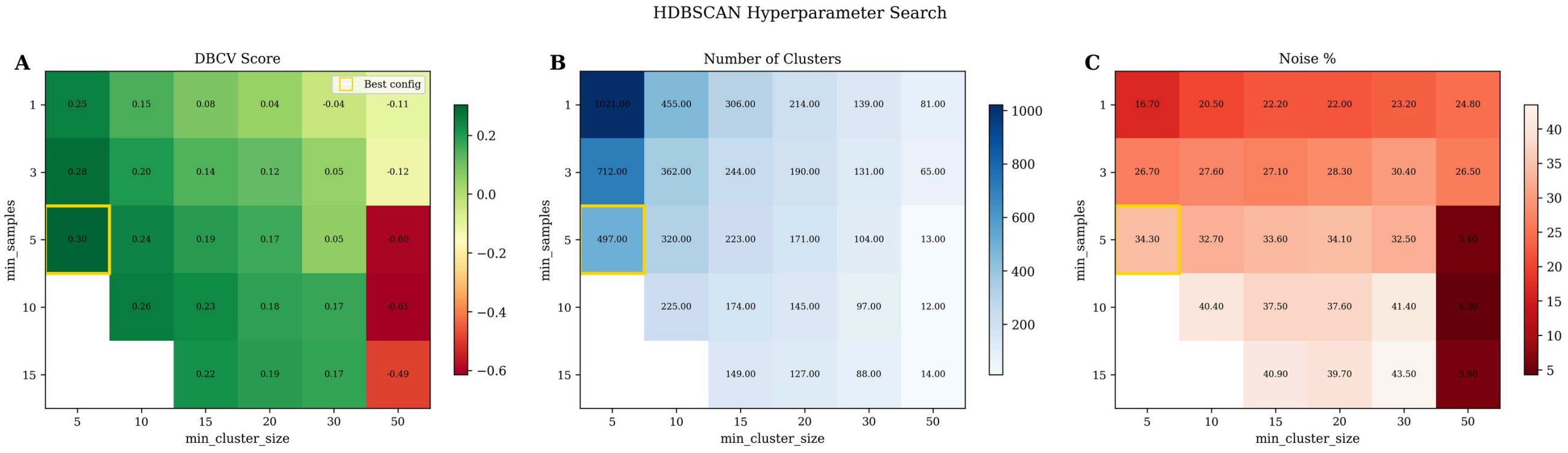